\documentclass[fleqn,usenatbib]{mnras}

\usepackage{newtxtext,newtxmath}

\usepackage[T1]{fontenc}

\DeclareRobustCommand{\VAN}[3]{#2}
\let\VANthebibliography\thebibliography
\def\thebibliography{\DeclareRobustCommand{\VAN}[3]{##3}\VANthebibliography}

\usepackage{graphicx}	
\usepackage{amsmath}	
\usepackage{lineo}

\graphicspath{{./}{figures/}}

\title[Moon-packing around an Earth-mass Planet]{Moon-packing around an Earth-mass Planet}

\author[S. Satyal et al.]{
Suman Satyal,$^{1}$\thanks{E-mail: suman.satyal@uta.edu}
Billy Quarles,$^{2,3}$
and Marialis Rosario-Franco$^{1,4}$
\\
$^{1}$Department of Physics, University of Texas at Arlington, Arlington, TX, 76019, USA\\
$^{2}$Department of Physics, Astronomy, Geosciences and Engineering Technology,
Valdosta State University,
Valdosta GA, 31698, USA\\
$^{3}$Center for Relativistic Astrophysics, School of Physics, Georgia Institute of Technology, Atlanta, GA 30332, USA \\
$^{4}$National Radio Astronomy Observatory,
Socorro NM 87801, USA
}

\date{Accepted 2022 July 28. Received 2022 July 27; in original form 2022 May 10}

\pubyear{2022}

\begin{document}
\label{firstpage}
\pagerange{\pageref{firstpage}--\pageref{lastpage}}
\maketitle

\begin{abstract}
All 4 giant planets in the Solar System host systems of multiple moons, whereas the terrestrial planets only host up to 2 moons. The Earth can capture small asteroids as temporary satellites, which begs the question as to how many moons could stably orbit the Earth, or an Earth-mass exoplanet. We perform a series of N-body simulations of closely-spaced equal mass moons in nested orbits around an Earth-mass planet orbiting a Sun-like star. The innermost moon begins near the host planet’s Roche radius, and the system is packed until the outermost moon begins near the stability limit for single moons. The initial spacing of the moons follows an iterative scheme commonly used for studies of compact planetary systems around single stars. For 3-moons system, we generate MEGNO maps to calculate periodic and chaotic regions and to identify the destabilizing MMRs. Our calculations show that the maximum number of moons depends on the assumed masses of the satellites (Ceres-, Pluto-, and Luna-mass) that could maintain stable orbits in a tightly-packed environment. Through our N-body simulations, we find stable configurations for up to $7\pm1$ Ceres-mass, $4\pm1$ Pluto-mass, and $3\pm1$ Luna-mass moons.  However, outward tidal migration will likely play a substantial role in the number of moons on stable orbits over the 10 Gyr stellar lifetime of a Sun-like star.

\end{abstract}




\section{Introduction} \label{sec:intro}

In our Solar System, most planets contain multiple satellites. Notably, the giant planets in the outer Solar System host multiple-moon systems. The only rocky planets that contain natural satellites are Earth and Mars; with $n_{moons} \le 2$. Given the discrepancies in total number of moons for the giant and terrestrial planets, it is expected that they experience different formation mechanisms and orbital evolution processes.

Several satellite formation theories have been proposed for both regular and irregular satellites around planets. The commonly accepted mechanism for regular satellite formation around giant planets is accretion from a circumplanetary disk. \citet{Canup2006} found a planet-disk mass ratio limit of $\sim 10^{-4}$; which is 2-3 orders of magnitude smaller than the mass fraction between the Earth and Moon. The results are consistent with simulations of disks with lesser initial mass that have resulted in formation of satellites that are not massive enough to clear out their orbits, but massive enough to start outward migration due to gravitational interaction with the disk \citep{Hyodo2015} and locked in mean motion resonance. 

Interactions between giant planets and circumplanetary disks heavily influence the processes of satellite formation, where the physical parameters that shape disk accretion and evolution have been constrained (\citet{Coradini2010}, and references therein). Specifically, the pressure and temperature profiles in the circumplanetary nebulae shaped the chemical gradients in the disk. These chemical gradients set the composition of the satellitesimals, which represent the building blocks of the present regular satellites. Additionally, further studies support the formation of natural satellites around Jupiter and Saturn within the framework of a quasi-steady state system \citep{Batygin2020}. A large-scale meridional flow of gas inside the planetary Hill sphere is developed in the later stages of planet formation which feeds a circumplanetary disk that expels gaseous material back into the parent nebula to maintain equilibrium in the system.

Recent studies have attempted to explain the origin of Galilean moon around Jupiter. \citet{Cilibrasi2021} studied less massive satellite systems through 3D radiative simulations and found that only $\sim 15\% $ of the resulting population is more massive than the Galilean satellites, which indicates low rates for tidal migrations and resonant captures are uncommon. \citet{Madeira2021} reproduced the system of Galilean satellites in a gaseous circumplanetary disc around Jupiter. However, their satellites have moderately eccentric orbits ($\sim0.1$), unlike the current real satellites. They propose a pre-existing resonance between Callisto and Ganymede that was broken over time via divergent migration due to tidal planet-satellite interactions. These same effects further damped the orbital eccentricities of these satellites down to their current values ($\lesssim 0.01$).

Satellites can be captured into the gravitational field of a planet resulting in a large semimajor axis, eccentricity, inclination, and in retrograde orbits. These irregular satellites offer important insight into the formation processes of regular satellites that likely formed in prograde rotating accretion disks. Irregular satellites can be captured by dissociation of a planetesimal binary in the planet's gravity field \citep{Vokrouhlicky2008}.  For example, Triton was likely captured by this process. \cite{Neto2006} shows that the satellites can be captured in prograde orbits (e.g., Leda, Himalia, Ysithea, and Elara by Jupiter) as a gas giant grows within the planetesimal disk.

Planet packing studies \citep{Smith2009,Quarles2018,Lissauer2021,Bartram2021} incorporate two or more Earth-mass planets orbiting a Sun-like star in low-eccentricity and low-inclination orbits. Two planets in circular orbits can be Hill stable if the fractional orbital separation is greater than 2.4($\mu_1$ + $\mu_2$)$^{1/3}$, where $\mu_1$, $\mu_2$ are the planets-Sun mass ratios \citep{Gladman1993}. Also, two small planets are stable if the initial semi major axis difference ($\Delta$) exceeds 2$\sqrt{3}$ mutual Hill radii, where systems of more than two planets are stable for $\Delta$ $\gtrsim$ 10 \citep{Chambers1996}. Similar studies of two-planet systems (equal mass, coplanar, circular orbits) exhibited stable chaos beyond the $2\sqrt{3}R_H$ separation \cite{Marzari2014}. Closely spaced five-planet systems can have shorter lifetimes when the planetary orbits begin with a non-zero initial eccentricity ($e_p = 0.05$) in contrast to initially circular orbits \citep{Gratia2021}.  Most $({\sim}72\%)$ closely packed five-planet systems with inclined orbits have a prompt collision after their first encounter, where a few $({\sim}1\%)$ survive up to $10^{7.5}$ orbits of the innermost planet \citep{Rice2018}.  \cite{Funk2010} studied hypothetical ultra-compact systems of up to 10 planets with Neptune-like masses (17 $M_\oplus$) within 0.26 AU and showed that some systems were stable even with a perturbing gas giant between $0.3-0.5$ AU.

The instability in the system arises because the energy and the angular momentum are not conserved due to the perturbations by the additional planet(s). The stability time varies linearly with the initial orbital spacing, and the stability time is significantly higher and can be packed twice as closely together for retrograde planets compared to prograde planets \citep{Smith2009}. On the other hand, planets in circumstellar orbits of binary system requires higher spacing than for the planets in single stars \citep{Quarles2018}. To further understand the dynamical stability of multiple moons, we follow the studies done on planet packing in circumstellar orbits in binary star system due to the similarities in orbital architectures (i.e., natural inner and outer boundaries).

 \citet{Domingos2006} derived a fitting formula for the stability limit of moons following previous studies of planet stability in binary star systems \citep{Rabl1988,Holman1999}.  The fitting formula defines a critical semi-major axis $a_c$ in units of the planetary Hill radius $R_{H,p}$ as $\sim 0.5\ R_{H,p}$ or $\sim 0.9\ R_{H,p}$ for satellites around giant planets in both prograde and retrograde orbits, respectively.  Eccentric orbits of either the planet or moon reduce these estimates in a nearly linear manner.  \cite{yagirl2020} clarified the stability limit of a moon in a prograde orbit as a fraction of the Hill radius (0.4$R_{H_p}$) through a series of N-body simulations that considered a wider range of initial mean anomaly for the satellite.  \citet{Quarles2021} revisited the stability limit for retrograde orbits where they showed a limit of $0.67 R_{H_p}$ and identified how the outer stability limit for a putative exomoon in $\alpha$ Centauri AB system varies due to a forced planetary eccentricity.

Earth has captured small bodies in temporary orbits, where these briefly captured rocks (quasi-satellites) either head into the atmosphere to become meteors (or meteoroids), or orbit the Earth until obtaining the necessary escape velocity to leave Earth's sphere of influence. The recently detected asteroid CD3, a quasi-satellite that remained in orbit for at least three years, is a prime example of this type of capture. \citet{Granvik2012} computed the natural Earth satellite capture probability from the near Earth object (NEO) population as a function of a NEO's heliocentric orbital elements. This numerical study included 10 million virtual asteroids and only 18,000 were captured in Earth orbit. They found that the average captured satellites make $\sim 3$ revolutions around Earth in 9 months.

The temporary orbits of quasi-satellites around Earth and that giant planets naturally contain multiple moons prompts a sensible question as to: "how many moons can stably orbit the Earth and how massive can they be?" In this paper, we perform a series of N-body simulations of closely-spaced equal-mass moons in nested orbits around an Earth-mass planet orbiting a Sun-like star to determine the maximum number of moons that could stably orbit the Earth and consider a range of three different prototype masses (Ceres-, Pluto-, and Luna-mass).  We use the term \emph{Luna} to identify a natural satellite that is similar in mass and radius to Earth's moon.

The methodology of our numerical simulations  are presented in Section \ref{section:theory} including the design of the system architectures to simulate the multiple moons (up to 9) in a Sun-Earth system. The results in Section \ref{section:results} consider Ceres-mass, Luna-mass, and Pluto-mass moons to identify the most stable orbital configuration for maximum number of moons orbiting Earth-mass planet . A summary of our results and a discussion of the broader context of multiple-moon system are in Section \ref{section:results} and \ref{section:conclusions}.

\section{Methodology}
\label{section:theory}

Earth and Mars are the only terrestrial planets with moons, where Earth hosts a single moon (the Moon or Luna) and Mars has two moons (Phobos and Deimos). Moon formation is a stochastic process, where the amount of material available largely dictates how many moons could form, but the goal of this work is to find the maximum number of moons that \emph{could} exist with respect to orbital stability constraints.

\subsection{Numerical Simulations using REBOUND}
We use the general N-body orbital evolution software \texttt{REBOUND} \citep{Rein2012} to examine the orbital stability of many equal-mass satellites orbiting an Earth-like planet, which in turn orbits a Sun-like star. \texttt{REBOUND} provides two algorithms (WHFast and IAS15) that are well-suited to evolve the hierarchical configuration of stars, planets, and moons. The accuracy of the numerical simulations using each algorithm is not substantially different when the initial timestep is set to 5\% of the innermost moon's initial orbital period and an 11th order symplectic corrector is used \citep{Rein2015} to minimize the energy error for WHFast. Therefore, we use the WHFast integrator for the sake of numerical expediency. 

Each simulation is evaluated up to $10^7$ times the period of the innermost moon $P_1$. The timescale for significant orbital evolution due to tides is much longer and thus we do not consider tidal effects in our N-body simulations.  Instead, we use a secular tidal model to evaluate the extent of the moons' outward migration.  An initial configuration is deemed stable, when all the moons are initially orbiting the host planet are present at the end of the simulation ($10^7$ $P_1$).  The dynamical timescale for moon systems is very short, where our timescale greatly exceeds the secular timescale for the Sun's forced eccentricity ($<100$ years; \cite{Andrade-Ines2017}).  As a result, systems far from stability boundaries will remain stable for billion-year timescales.  Indeed, our own Moon will evolve onto an unstable orbit \citep{Sasaki2012} eventually, but this timescale is longer than the main sequence lifetime of the Sun and renders the issue of stability moot due to the possible engulfment of the Earth-Moon system.

Unstable initial conditions are those that result in a close approach (within the Roche radius) with the host planet, collisions between neighboring moons or when a moon's apocenter extends beyond the outer stability limit measured in terms of the planet's Hill sphere ($Q_{sat} > 0.4R_{H,p}$;  \citet{yagirl2020}).  Although collisions are possible in our simulations, they are rare and scattering events that transport a moon beyond the outer stability limit represent the vast majority of outcomes. An individual simulation is terminated once an instability occurs, which represents the simulation lifetime $t_{sim}$.  We scale the simulation lifetime by the orbital period of the innermost moon $T_1$ to obtain the number of orbits completed by the innermost moon, $N_1 = t_{sim}/T_1$.  In all of the simulations, the host planet begins on an orbit that is identical to the Sun-Earth system using the JPL Horizons lookup feature of \texttt{Rebound} so that $a_p \approx 0.999$ AU and $e_p \approx 0.0167$.  As a result, the Sun will perturb each moon's orbit and lead to a small forced eccentricity \citep{Andrade-Ines2017,Quarles2021}.  

\subsection{System Architecture and Formulation of Orbital Spacing}
\label{sec:architecture}

Although we neglect the long-term orbital effects of tides, tidal forces do place a lower limit on how close a smaller body (e.g., planet or moon) could orbit its parent body (e.g., star or planet). Interior to the Roche limit, the orbiting body gets disintegrated by the tidal force when it overcomes the surface gravity. For a moon with a mass $m_{sat}$ and radius $r_{sat}$ orbiting a planet with mass $m_p$ and radius $r_p$, the Roche radius (via the fluid definition) is given as,
\begin{equation} \label{eqn:Roche1}
    R_{Roche} \approx 2.44 ({m_p/m_{sat}})^{1/3}r_{sat},
\end{equation}
or 
\begin{equation}\label{eqn:Roche2}
    R_{Roche} \approx 2.44 ({\rho_p/\rho_{sat}})^{1/3}r_p,
\end{equation}
which depends on the bulk density of planet $\rho_p$ and moon $\rho_{sat}$ through $m_p/m_{sat} = (\rho_{sat}/\rho_p)(r_p/r_{sat})^3 $.

In the three-body problem, the Hill sphere (or radius) defines a region of space where a planet's gravity dominates over the host star's pull. To first approximation in the planetary eccentricity, the Hill radius for a moon is truncated by the host planet's pericenter by,

\begin{equation} \label{eqn:Hill_radius}
    R_{H,p} = a_p(1-e_p)\left(\frac{m_p}{3M_\star}\right)^{1/3},
\end{equation}
which depends on the planet's semimajor axis $a_p$, eccentricity $e_p$, mass $m_p$ and the host star's mass $M_\star$. In the case of large moons, Equation \ref{eqn:Hill_radius} requires modification by replacing the planet mass with $m_p^\prime$, which is the total mass of $i$ moons added to the planet mass (i.e., $m_p^\prime = m_p + i m_{sat}$). The Hill radius is a theoretical point of stability at an instant in time, where many numerical simulations have shown that the outer stability limit actually lies within about half of the Hill radius \citep{Domingos2006, yagirl2020}.

Each of the moons begin on a circular, coplanar orbit around an Earth-like planet. The initial semimajor axis of each moon is determined by a unit-less spacing parameter $\beta$ for each simulation. A similar procedure has been used for the study of planet packing around single stars \citep{Chambers1996,Smith2009,Obertas2017} and in binary star systems \citep{Quarles2018}.  The spacing parameter $\beta$ is a normalized separation between the two nearby orbits and uses the \emph{mutual Hill radius} $R_{H,m}$ between adjacent moons for the normalization. The mutual Hill radius calculated using the total mass that lies interior to ith moon is given by $\widetilde{M}_i = m_p + (i-1)m_{sat,i}$. The mutual Hill radius for two consecutive moons with mass $m_i$ and $m_{i+1}$ is defined by:

\begin{equation}\label{eqn:Mutual_Hill}
    R_{H,m} = (a_i + a_{i+1})X \;\; {\rm and} \;\; X = \frac{1}{2}\left[\frac{m_i+m_{i+1}}{3\widetilde{M}_i}\right]^{1/3},
\end{equation}
where a recurrence relation defines the semimajor axis of the $i+1$-th outer moon using the semimajor axis of the inner moon $a_j$, the input parameter $\beta$, and $X$ (from Equation \ref{eqn:Mutual_Hill}) as
\begin{equation}\label{eqn:semi}
    a_{i+1} = a_i\left(\frac{1+\beta X}{1-\beta X}\right).
\end{equation}

The above equations can be used to describe the mutual Hill radius for a wide range of orbital architectures.  In our problem the $m_i + m_{i+1}$ factor can be replaced with $2m_{sat}$ because we use equal-mass moons.

The initial system setup includes the Sun, Earth and a single moon with its semimajor axis $a_1$ at 2 $R_{Roche}$. The position of the subsequent moons (up to 9) are then prescribed using Equation \ref{eqn:semi} for a chosen value of $\beta$. Following \cite{Smith2009} and \cite{Quarles2018}, the initial mean anomaly is set by using multiples of the golden ratio through $2\pi i \phi$ radians = $360i\, \phi$ degrees, where the golden ratio is $\phi = (1 + \sqrt{5})/2$. Using the golden ratio in this context allows us to add the moons to the system so that no pair of moons is initialized at conjunction. This also helps to avoid the mean motion resonances (MMRs) between moons because the first and second order MMRs can reduce a system's lifetime \citep{Quarles2018}.

\begin{figure*}
  \includegraphics[scale=.60]{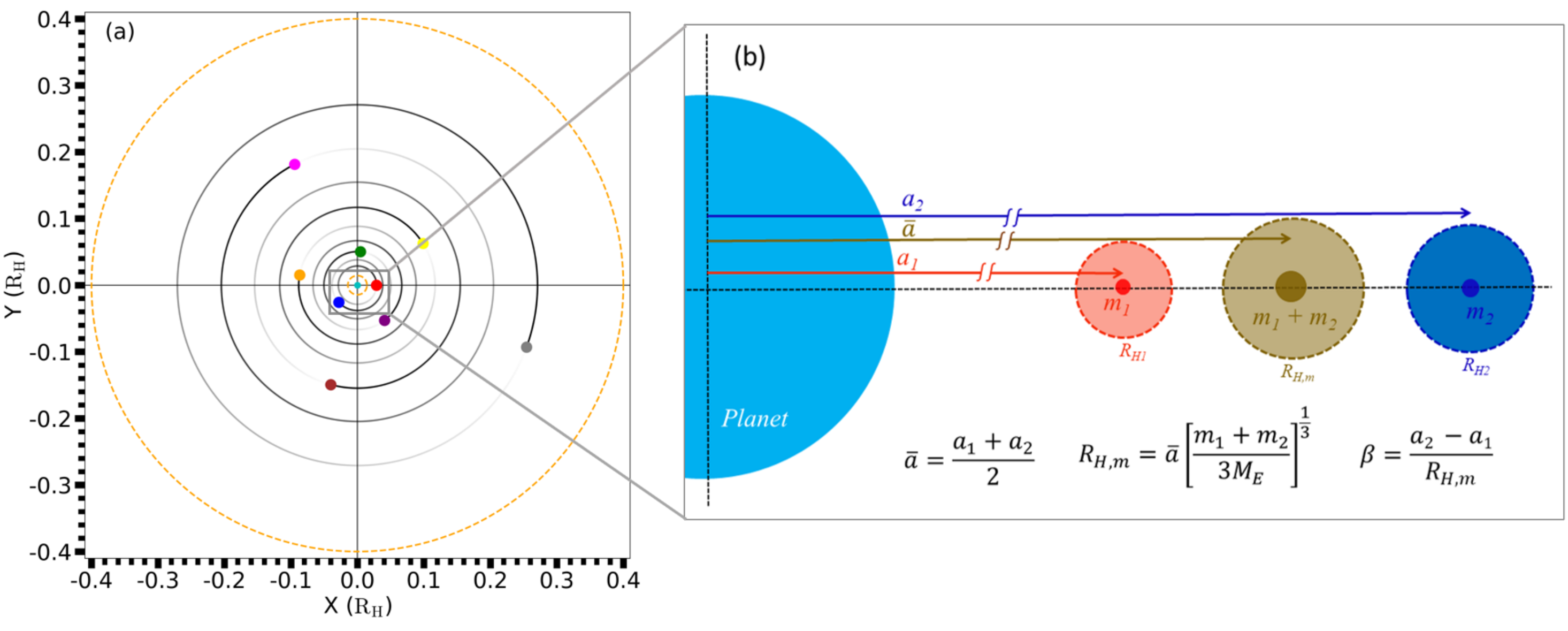}
    \caption{(a) Schematic of an example orbital architecture showing the initial positions of Ceres-mass moons orbiting an Earth-mass planet. The orbital sizes are determined using the orbital spacing parameter, $\beta$ described in Section \ref{sec:architecture}, where $\beta$  = 6. The innermost and outermost dotted circular lines (in orange) denote the inner (Roche radius) and outer stability limit (0.4 $R_H$). The satellites are color-coded (red, blue, green, .. , gray) from the innermost to outermost moon, where this color scheme is consistent throughout the paper. The moons' sizes are not to scale. (b) Schematic showing how the spacing parameter $\beta$ is calculated for the two innermost moons (red and blue). The brown body represents the mutual Hill radius. The Hill radius for $m_1$, $m_2$, ($m_1$+$m_2$), and radius of the planet are drawn to scale. The semi-major axes ($a_1$, $a_2$ and $\bar{a}$) have the same units as the Hill radii but their scale is broken to minimize the plotting area.}
\label{fig:1}
\end{figure*}

Figure \ref{fig:1} illustrates the orbital architecture for 8  Ceres-mass moons, including their initial angular position (in Fig. 1a). The color scheme for the moon index is consistently used through out this paper. In this color scheme, the Earth is represented by cyan and the orbital dots (red, blue, green, ..., gray) denote each moon in the order from inner to outer. The time variation of a particular moon follows the same color-code in later sections. The inner boundary using Earth's Roche radius $R_{Roche}$ and outer boundary using the stability limit for a single moon ($0.4R_H$) are indicated by the dashed orange circles, respectively. The axes' units in Fig. \ref{fig:1} are converted to Hill radius to provide a logical representation of the Earth's Hill sphere. Since the Hill radius of any outer moon (compared to an immediate inner moon) is longer, the orbital spacing increases going from innermost to outermost orbit by a factor of $\sim$1/3. The displayed orbital spacing and the number of orbits between the stability regions is calculated assuming $\beta$  = 6.

Figure \ref{fig:1}b is the projection of the Earth and 2 innermost moons that are drawn within the gray box in Fig. \ref{fig:1}a. This schematic highlights geometrically how the spacing parameter $\beta$ distributes two consecutive moons. The Hill radius of each moon individually ($R_{H1}$ and $R_{H2}$), and their mutual Hill radius ($R_{H,m}$) scaled by the planet's radius $R_p$. The semimajor axis of each moon has a scale break so that the all the bodies can fit on the page. The central body (brown) represents the total moon mass ($m_1$ +$m_2$) with the mutual Hill radius $R_{H,m}$. A similar setup is employed when we consider larger masses for the moons (Pluto-mass and Luna-mass).  Note that the mutual Hill radius scales with the assumed mass of the moons, where increasing the moons from a Ceres-mass to a Pluto-mass also increases the mutual Hill radius by a factor of $(m_{Pluto}/m_{Ceres})^{1/3} \approx 2.4$.  The mutual Hill radius between Luna-mass moons is $\sim4.3\times$ larger than Ceres-mass moons.  As a result, more massive moons will necessarily be limited to smaller values of $\beta$ so that the outermost moon doesn't exceed the outer stability limit.

\subsection{Initial System Parameters}
\label{sec:parameters}
The number of moons that an Earth-like planet can host depends on the assumed satellite masses and their spacing that modulate the gravitational interactions between the satellites.  We use three categories (Ceres, Pluto, or Luna ) as prototypes for different sized moons in terms of their mass and radius. The satellite prototypes are used because they represent the most massive object in the asteroid belt (Ceres) and the most massive object in the Kuiper belt (Pluto). The Moon (Luna) is used due to its large relative mass/size among the natural satellites.  Phase lag, or constant Q, tidal models suggest that more massive satellites than those we consider can escape an Earth-like planet through outward tidal migration \citep[][see their Fig. 9]{Quarles2021} and thus we limit our study to planet-satellite mass ratios $\lesssim 0.02$.  Probing to smaller masses runs into problems, where we need to consider non-gravitational effects (e.g., Yarkovsky effect or Poynting-Roberson drag). Our study is limited to large and massive objects so that such effects are negligible and can be ignored.

The mean density for all three bodies varies which results a difference in the Roche radius and the semimajor axis of innermost satellite. Table \ref{tab:Orb_Param} provides the initial values used to define the Roche radius that will scale the innermost satellite orbit. Starting the innermost satellite at the Roche radius could bias our results when that satellite's eccentricity evolves and its pericenter is raised (i.e., $q_1<R_{Roche}$).  We begin the innermost satellite on an initially coplanar, circular orbit with a semimajor axis that is 2 times the Roche radius ($a_1 = 2R_{Roche}$). 

To define the first trial value $\beta_{min}$ in the orbital spacing of the satellites, we use the dynamical results from previous planet packing studies \citep[e.g.,][]{Gladman1993,Chambers1996,Smith2009,Obertas2017,Quarles2018} that show a minimum spacing ($\beta_{min}=2\sqrt{3}$), where smaller values are unstable (and chaotic) due to the overlap of first-order MMRs \citep{Wisdom1980}.  For $\beta\gtrsim\beta_{min}$, there is an expected transition regime up to a critical value $\beta_{crit}$ that represents a broad plateau of stable configurations \citep{Obertas2017,Lissauer2021}. The extent of the plateau is expected to change slightly beyond our integration timescale, especially near the MMRs, inner most stable beta, and outermost stable beta due to stochastic encounters.  However, this does not exclude the existence of stable conditions within such plateaus. \cite{Quarles2018} showed for binary systems that a maximum value $\beta_{max}$ signifies another transition regime, but from stable to unstable due to MMRs with an external perturber.  We adapt the results from \cite{Quarles2018} (see their Eqn. 3) to calculate the maximum $\beta$ through the following:

\begin{equation} \label{eqn:beta_max}
     \beta_{max} = \left(\frac{(a_N/a_1)^{\frac{1}{N-1}}-1}{(a_N/a_1)^{\frac{1}{N-1}}+1}\right) \left(\frac{12m_p}{m_i}\right)^{1/3},
\end{equation}

which depends on the number of moons N, the innermost orbit $a_1$, the outermost orbit $a_N$, the planet's mass $m_p$, and each satellite's mass $m_{sat}$.  For satellite systems with small $N$, the value of $\beta_{max}$ can be large and thus, we only iterate up to $\beta = 10$.  As the number of moons $N$ increases, we can use Eqn. \ref{eqn:beta_max} to verify whether a instability transition occurs.  Each of our satellite prototypes (Ceres, Pluto, or Luna) are varied in the initial $\beta$ starting from $2\sqrt{3}\approx 3.5$ up to 10 through steps of 0.01.

The spacing parameter $\beta$ increases the semimajor axis of subsequent moons and the associated orbital periods.  As a result, the ratio of orbital periods between a pair of moons can start as a near integer ratio to form an MMR.  Although we take steps to avoid MMRs through the initial phasing of the moons, some configurations can still enter into the MMR at least temporarily.  To identify the expected locations of MMRs as a function of $\beta$ \citep{Obertas2017}, we use the semimajor axis ratio $a_{i+1}/a_i$ from Eqn. \ref{eqn:semi} and apply Kepler's 3rd law to get

\begin{equation} \label{eqn:MMR}
    \frac{P_{i+1}}{P_i} = \left(\frac{1+\beta X}{1-\beta X}\right)^{3/2},
\end{equation}

for an adjacent pair of moons.  Equation \ref{eqn:MMR} provides the expected location of MMRs assuming that the Sun has a negligible influence on the moons and can be generalize to any pair of moons following the formalism for planet packing \citep{Obertas2017}.

\begin{table}
\centering
\caption{Initial satellite parameters (mass, radius, and density) that define the innermost orbit $a_1$ in terms of $R_{Roche}$ (see Eqns. \ref{eqn:Roche1} and \ref{eqn:Roche2}) for our $N$-body simulations. The volumetric mean radius of each satellite type is used, where $r_p = 6371$ km for the Earth.  The period is provided for easier comparisons with other known satellite systems. \label{tab:Orb_Param}}

\begin{tabular}{lcccccc}
\noalign{\smallskip}
\hline
\noalign{\smallskip}
Body     & Mass                   & Radius & Density & $a_1$ & $a_1$ & Period\\
         & (M$_\oplus$)            & (km)   & (g/cm$^3$) & (au) & (R$_\oplus$) & (days)\\
\noalign{\smallskip}
\hline
\noalign{\smallskip}
Ceres   & 0.00015  & 469.7  & 2.162 & 0.000288 & 6.75498 & 1.010\\
Pluto   & 0.0022   & 1188   & 1.854 & 0.000299 & 7.01298 & 1.090\\
Luna    & 0.0123   & 1737.4 & 3.344  & 0.000247 & 5.79333 & 0.807\\
\noalign{\smallskip}
\hline
\noalign{\smallskip}
\end{tabular}
\end{table}

\begin{table}
\centering
\caption{Values of $\beta_{max}$ (see Eqn. \ref{eqn:beta_max}) when varying the number of moons $n$ using the innermost orbit $a_1$ and mass $m_j$ for each moon type (see Tab. \ref{tab:Orb_Param}).  The red text marks when $\beta_{max}<2\sqrt{3}$. \label{tab:beta_max} }

\begin{tabular}{lccc}
\noalign{\smallskip}
\hline
\noalign{\smallskip}
 $N$     & \multicolumn{3}{c}{$\beta_{\rm max}$}\\
       & $m_{\rm Ceres}$    & $m_{\rm Pluto}$   & $m_{\rm Luna}$\\
\noalign{\smallskip}
\hline
\noalign{\smallskip}
3	&	24.85	&	10.04	&	5.96	 \\
4	&	17.76	&	7.17	&	4.29	 \\
5	&	13.68	&	5.51	&	\textcolor{red}{3.31}	 \\
6	&	11.08	&	4.46	&	\textcolor{red}{2.69}	 \\
7	&	9.30	&	3.75	&	\textcolor{red}{2.26}	 \\
8	&	8.00	&	\textcolor{red}{3.22}	&	\textcolor{red}{1.94}	 \\
9	&	7.02	&	\textcolor{red}{2.83}	&	\textcolor{red}{1.71}	 \\
10	&	6.25	&	\textcolor{red}{2.52}	&	\textcolor{red}{1.52}	 \\
\noalign{\smallskip}
\hline
\noalign{\smallskip}
\end{tabular}
\end{table}

\section{Results and Analysis}
\label{section:results}

\subsection{Case Study 1: Ceres-mass Moons}
\label{ceres}
Using Ceres-mass moons, we perform numerical simulations varying the number of moons ($n=3-9$) and their orbital spacing through the spacing parameter ($2\sqrt{3}\le \beta \le 10$).  Each of the simulations are integrated up to $10^7$ orbits of the innermost moon, which begins at twice the Roche radius.  Planet packing studies using a single star \citep{Smith2009,Obertas2017} showed that stability is attained for 3-5 planets when $\beta \sim 7-10$, where the theoretical minimum stable value $\beta_{min}$ is $2\sqrt{3}$ from Hill stability \citep{Gladman1993}.  Table \ref{tab:beta_max} provides estimates for the maximum spacing $\beta_{max}$ (see Eqn. \ref{eqn:beta_max}) for each moon prototype. If we assume a maximum spacing equal to the minimum value for stability in single star systems ($\beta \sim 7$), then we would estimate (using Tab. \ref{tab:beta_max}) a maximum of 9 Ceres-mass moons that \emph{could} be stable.  However, we confirm this estimate by performing simulations with 10 moons and find that all the simulations are short-lived. Figure \ref{fig:1}a demonstrates how much of the parameter space is filled, where 9 Ceres-mass moons with $\beta = 6$ nearly reaches the outer stability (orange dotted circle).

\begin{figure*}
    \includegraphics[width=0.7\linewidth]{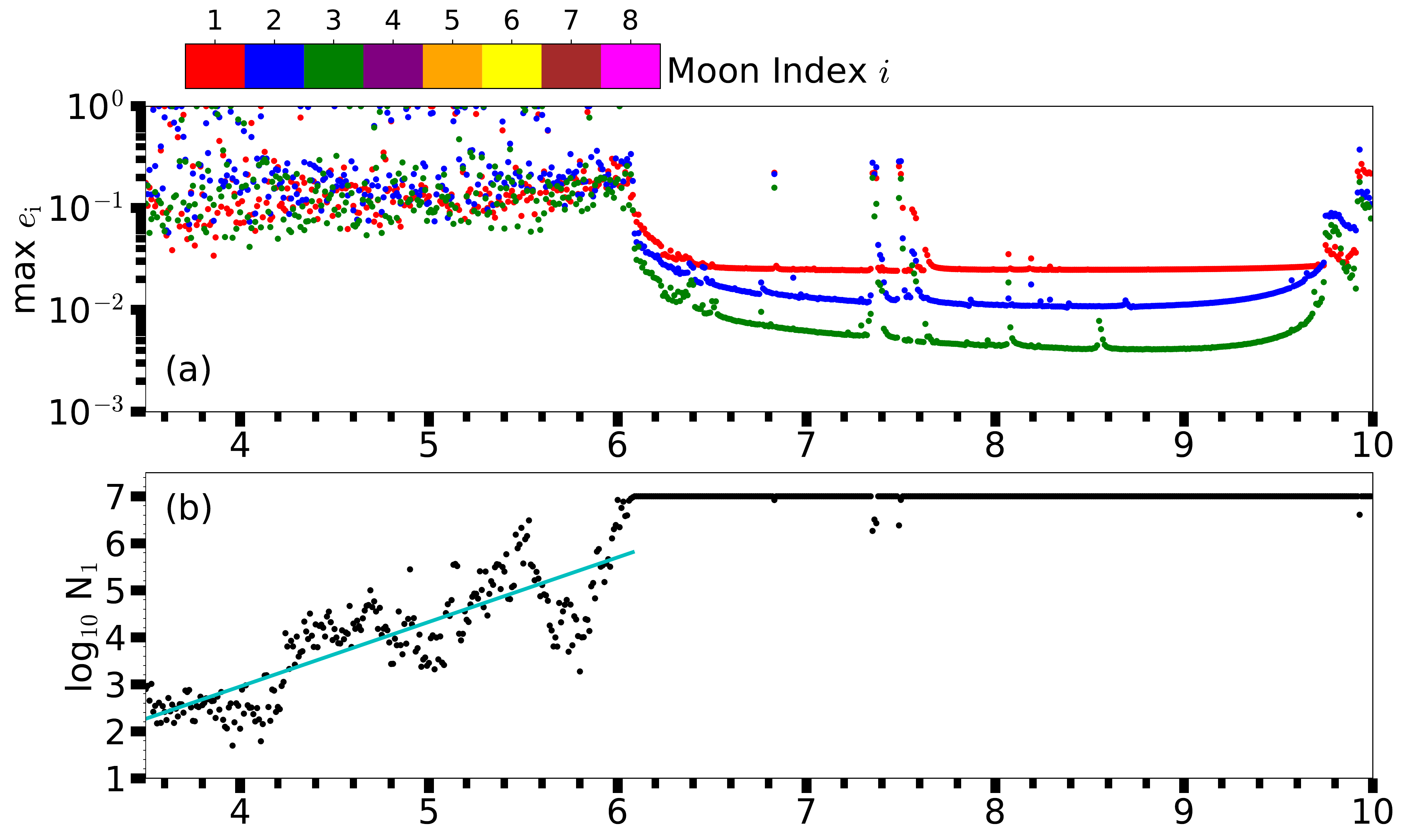}\\
    \includegraphics[width=0.7\linewidth]{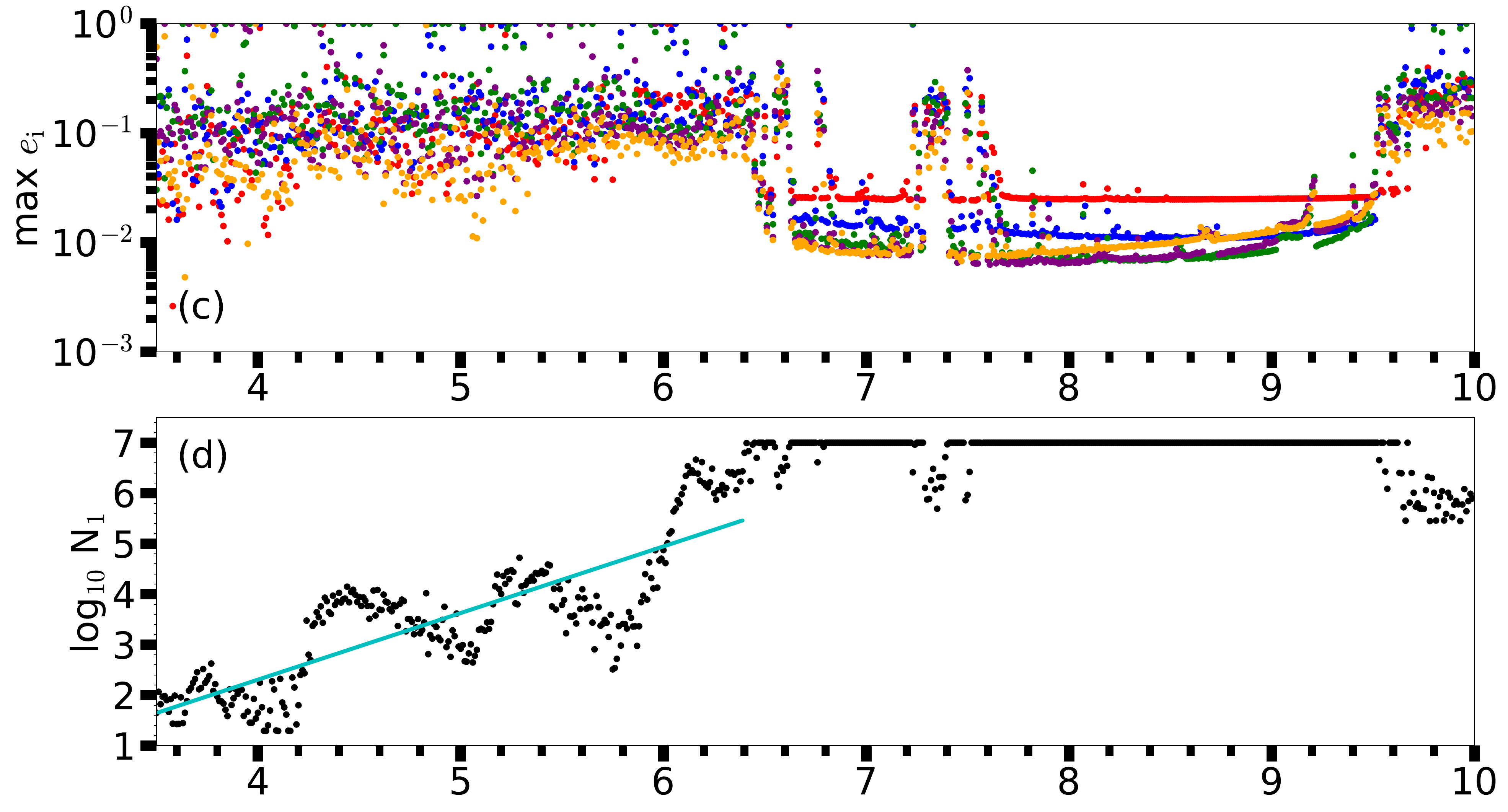}\\
    \includegraphics[width=0.7\linewidth]{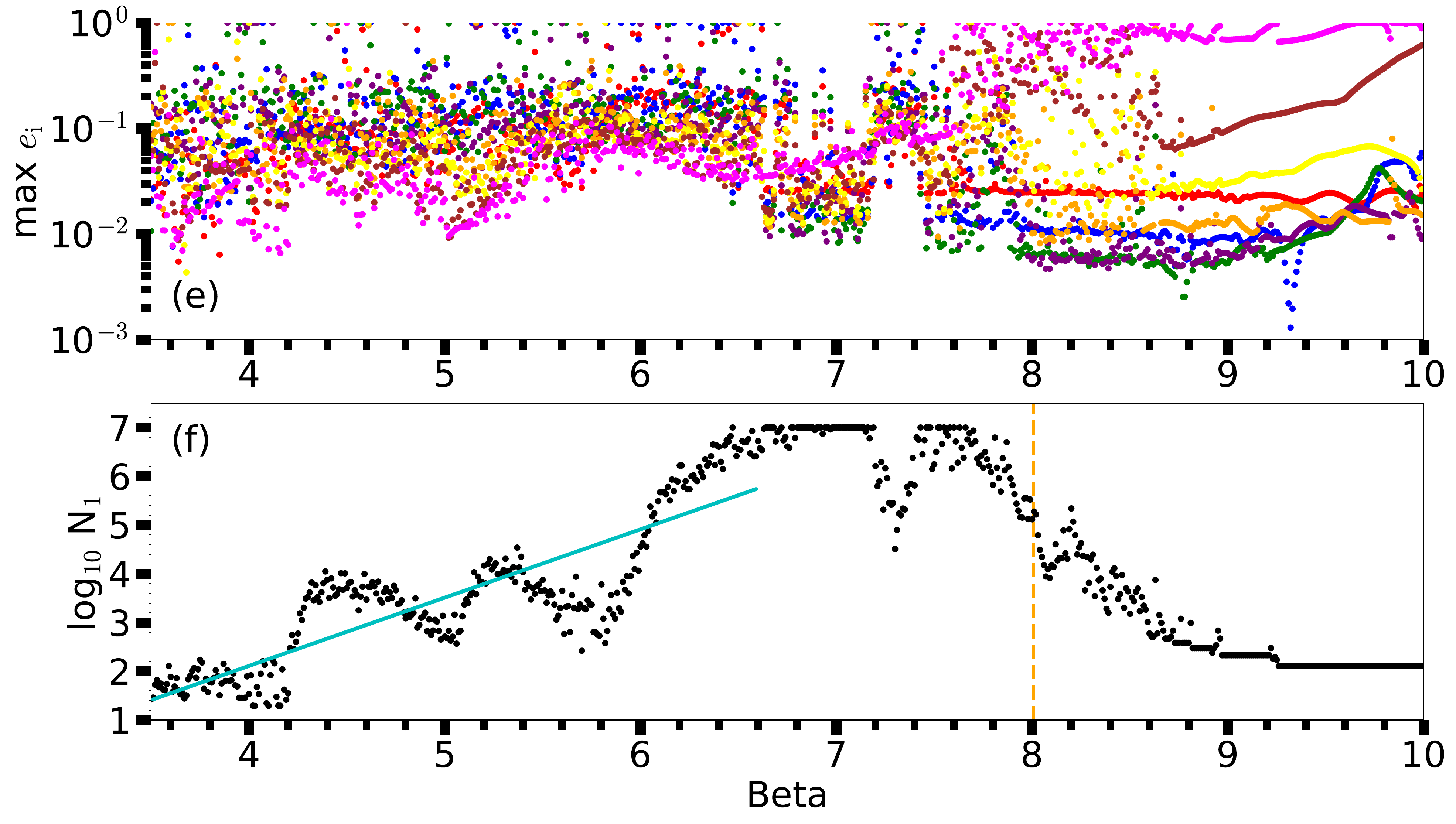}\\
\caption{Maximum eccentricity (color-coded) attained for an initial spacing parameter $\beta$ for 3 (panel a), 5 (panel c), and 8 (panel e) Ceres-Mass moons. The system lifetime measured in terms of the number of orbits $N_1$ completed by the innermost moon is plotted on a logarithmic scale for 3 (panel b), 5 (panel d), and 8 (panel f) Ceres-Mass moons. The orange dashed-line marks the maximum $\beta$ for highest number of moons (see Table \ref{tab:beta_max}). The cyan line is the log-linear fit to the unstable lifetimes and is discussed in Sec. \ref{sec:log-linear}.}
\label{fig:2}
\end{figure*}

In addition to the simulation lifetime (scaled by $T_1$), we track each satellite's maximum eccentricity ${\rm max}\:e_i$.  For most of our simulations, each satellite begins on a circular orbit and thus, the minimum eccentricity is zero.  Figure \ref{fig:2} shows the maximum eccentricity attained by each moon (Fig. \ref{fig:2}a) and the simulation lifetime (Fig. \ref{fig:2}b) with respect to the initial orbital spacing parameter $\beta$ for 3 Ceres-mass moons.  Figures \ref{fig:2}c and \ref{fig:2}d are similar to \ref{fig:2}a and \ref{fig:2}b, but evaluate 5 moons. Similarly, Figs. \ref{fig:2}e and \ref{fig:2}f evaluate 8 moons. The evaluation of 9 moons is not shown here as the system was unstable in less than a million years. The maximum eccentricity is color-coded (top of the figure) and indicates the index of the moon, where $i=1$ and $i=8$ refer to the innermost and outermost moons, respectively.

Since the Earth-mass host planet begins with a non-zero eccentricity ($e_p \approx 0.0167$), then each satellite experiences a forced eccentricity, which arises in solving the secular quadrupole problem of the orbit-averaged disturbing function \citep{Andrade-Ines2017}.  The magnitude of the forced eccentricity due to the Sun is typically small, but the eccentricity growth from moon-moon interactions can quickly drive up a moon's eccentricity enough ($e_i \sim 0.1$) for orbit crossings to occur.  In Fig. \ref{fig:2}a, the maximum eccentricity of each moon $e_i$ is high due to the chaotic overlap of first order mean motion resonances (MMRs; \cite{Deck2013}) for $2\sqrt{3}\le \beta \lesssim 6$.  Once $\beta>6$, the gravitational perturbations from moon-moon interactions weakens and the moon pairs exit the chaotic zone.  Beyond $\beta \sim 6$, the maximum eccentricity of each moon remains low ($\mathbf{\lesssim 0.1}$), but non-zero.  There are spikes in the maximum eccentricity (in Fig. \ref{fig:2}a) that correlate with dips in the lifetimes $(\log_{10}\,N_1)$ measured in Fig. \ref{fig:2}b.  These anomalous features ($\beta = 7.2,\,7.4,\,7.6,\,\ldots$) correspond to first order MMRs between adjacent pairs of moons.  The rise in maximum eccentricity at $\beta\sim 9.8$ corresponds to the 2:1 MMR, which is expected at $\beta \approx 9.78$ \citep{Obertas2017}.  The stable plateau for three Ceres-mass moons continues until $\beta \sim 24$, where the apocenter of the outermost moon will extend beyond the outer stability limit. 

Considering systems with additional satellites modifies Figs. \ref{fig:2}a and \ref{fig:2}b by lowering $\beta_{\rm max}$ (see Table \ref{tab:beta_max}) and the limit on the number of moons is restricted by the minimum value, $2\sqrt{3}$.  The maximum eccentricity for systems of five and eight Ceres-mass moons are given in Figs. \ref{fig:2}c and \ref{fig:2}e, respectively, while the lifetimes are shown in Figs. \ref{fig:2}d and \ref{fig:2}f, respectively.  These panels illustrate similar features as Figs. \ref{fig:2}a and \ref{fig:2}b, where the tail of stability is apparent at $\beta \approx 8$ (see vertical dashed line) Fig.  \ref{fig:2}f.  For $6.6\lesssim \beta \lesssim 7.6$, a system of eight moons is on the border of stability as evidenced by the growth of eccentricity with $\beta$ for the outermost moon and additional effects (e.g., tides) may prevent long-term stability in general.  A dip appears at $\beta \sim 7.3$ in Fig. \ref{fig:2}f, that corresponds to the 5:3 MMR between adjacent moons.  The width of the MMRs grows as moons are added to the system because there are more small perturbations possible that can push a moon into a MMR as the system evolves (i.e., evolution of a moon's semimajor axis or $\beta$).

Tracking the maximum eccentricity of each moon's orbit shows which values of $\beta$ are likely to produce unstable systems.  The timescale for $10^7$ orbits of the innermost moon is only $\sim20,000$ yr, where longer simulations are impractical due to the small timestep required for accurately evolving each system.  The gravitational interactions between moons also occur on a short timescale, which further necessitates a small integration timestep.  Despite these limitations, the maximum eccentricity provides a good heuristic to show the likely long-term stability of moons.  Although not shown here, we perform simulations with 4, 6, and 7 moons and confirm the apparent trends in stability.

\subsection{Case Study 2: Pluto-mass Moons} \label{pluto}
 The mutual Hill radius increases with Pluto-mass moons, which corresponds to a larger physical spacing between moons through the parameter $\beta$.  The expected locations of the MMRs between moons more strongly depend on $\beta$, where they move to lower values by a factor of $\sim2.44$, or $\left(m_{Pluto}/m_{Ceres}\right)^{1/3}$.  With these expectations, we perform simulations of Pluto-mass moons following the same procedure from Sect. \ref{ceres}.

\begin{figure*}
  \includegraphics[width=0.7\linewidth]{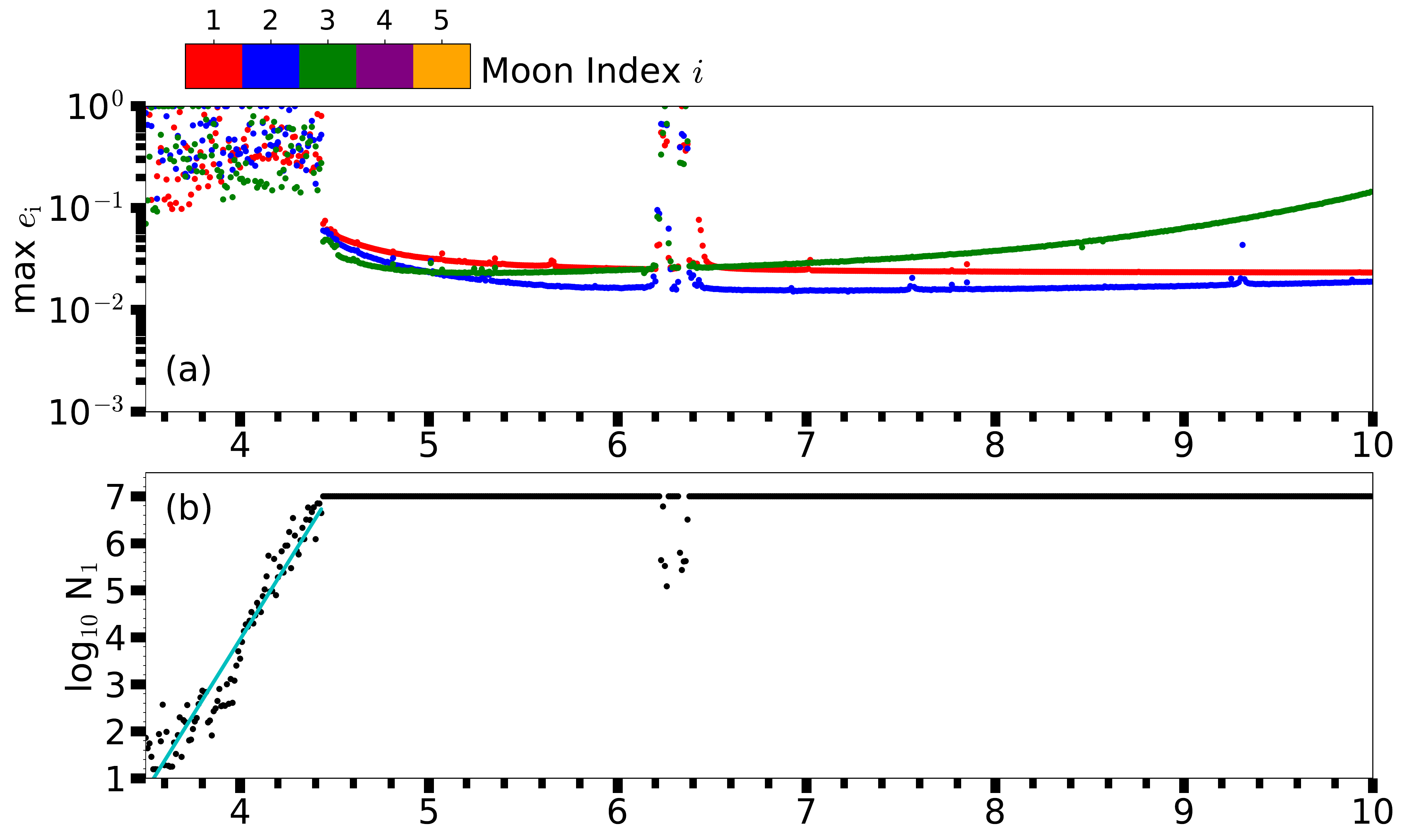}\\
  \includegraphics[width=0.7\linewidth]{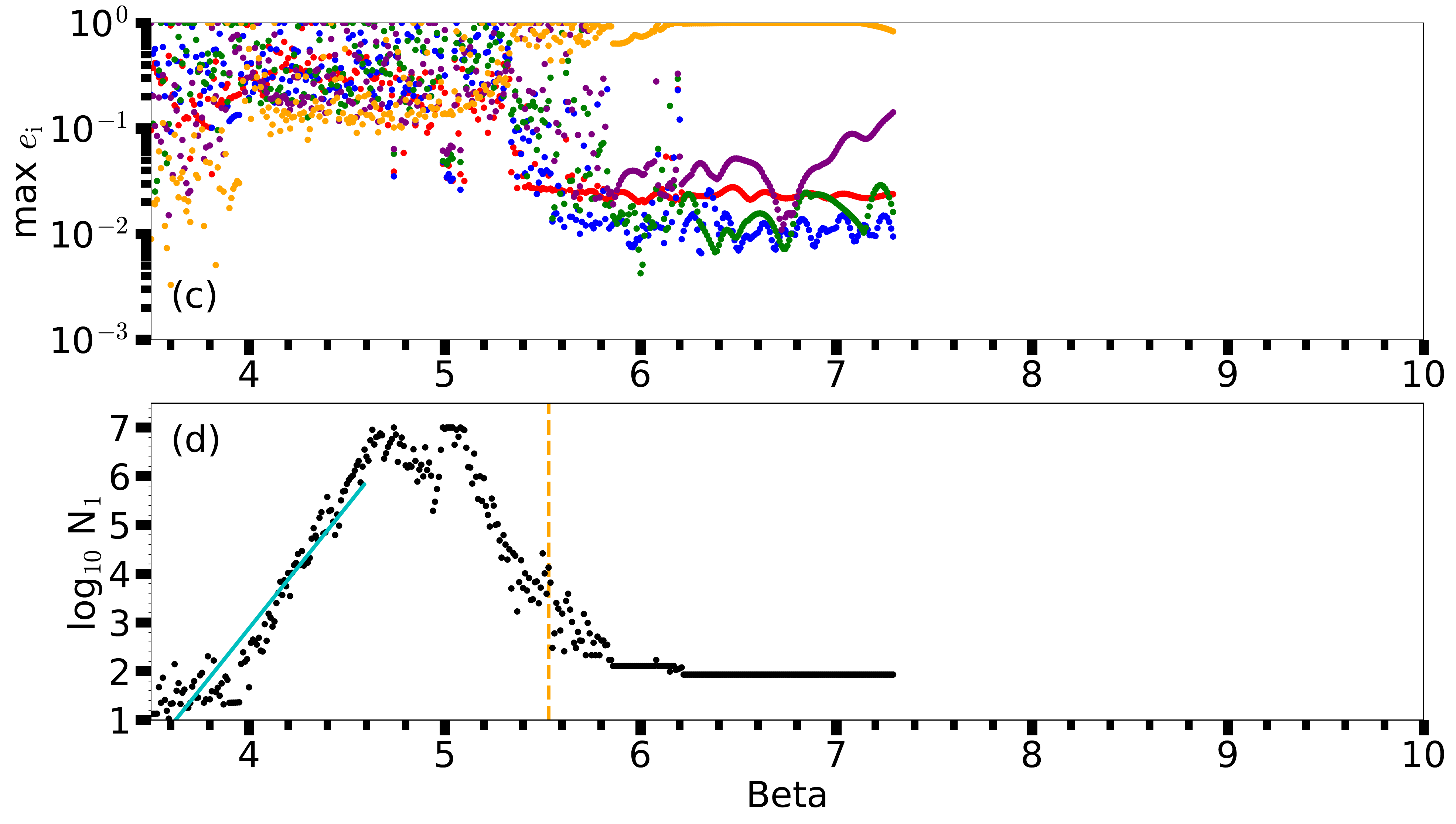}\\
\caption{Similar to Fig. \ref{fig:2} but for 3 (panels a \& b) and 5 (panels c \& d) Pluto-Mass moons.}
\label{fig:3}
\end{figure*}

Figure \ref{fig:3} illustrates how the maximum eccentricity of each moon and system lifetime varies with respect to the spacing parameter $\beta$ for a system of 3 (Figs. \ref{fig:3}a and \ref{fig:3}b) and 5 (Figs. \ref{fig:3}c and \ref{fig:3}d) moons.  From Table \ref{tab:beta_max}, we expect a system of 3 moons to be stable up to $\beta \sim 10$, where Fig. \ref{fig:3}a shows the increasing maximum eccentricity attained by the outermost moon $(i=3)$ as $\beta$ increases.  The minimum spacing appears at $\beta \sim 4.5$ because the 2:1 MMR is expected at $\beta \approx 4$ and planet packing studies have shown broad stability occurring beyond this MMR \citep{Smith2009,Obertas2017,Quarles2018,Lissauer2021}. The underlying physical cause is the wider separation of libration zones for the first order MMRs and the wider physical separation (i.e., less gravitational perturbations) between moons. The spike in Fig. \ref{fig:3}a and corresponding dip in \ref{fig:3}b at $\beta$=6.3 shows the location of the 3:1 MMR, where instabilities occur over longer timescales.

For five Pluto-mass moons (in Figs. \ref{fig:3}c and \ref{fig:3}d), there are very narrow ranges $(4.8 \lesssim \beta \lesssim 5.1)$ for which the system can survive up to $10^7$ orbits of the innermost moon.  As shown earlier for a system of three moons, the 2:1 MMR delineates a lower boundary and puts a limit on stability at $\beta \sim 4.5$, but the outermost moon's eccentricity grows to nearly 1 at $\beta \sim 5.5$, where the perturbations from the Sun are driving its eccentricity to such high values.  From Table \ref{tab:beta_max}, the outermost moon begins beyond the outer stability limit \citep{yagirl2020} at $\beta \sim 5.5$ (yellow dashed line in Fig. \ref{fig:3}d).  In between these boundary conditions (MMR overlap and perturbation from the Sun), a system of five Pluto-mass moons could be stable if the apsidal precession rate of the moons is similar enough to prevent orbital overlap or the moons remain out-of-phase in an MMR to avoid collision (e.g., the 3:2 resonance between Neptune and Pluto).  

\subsection{Case Study 3: Luna-mass Moons}
\begin{figure*}
  \includegraphics[width=0.7\linewidth]{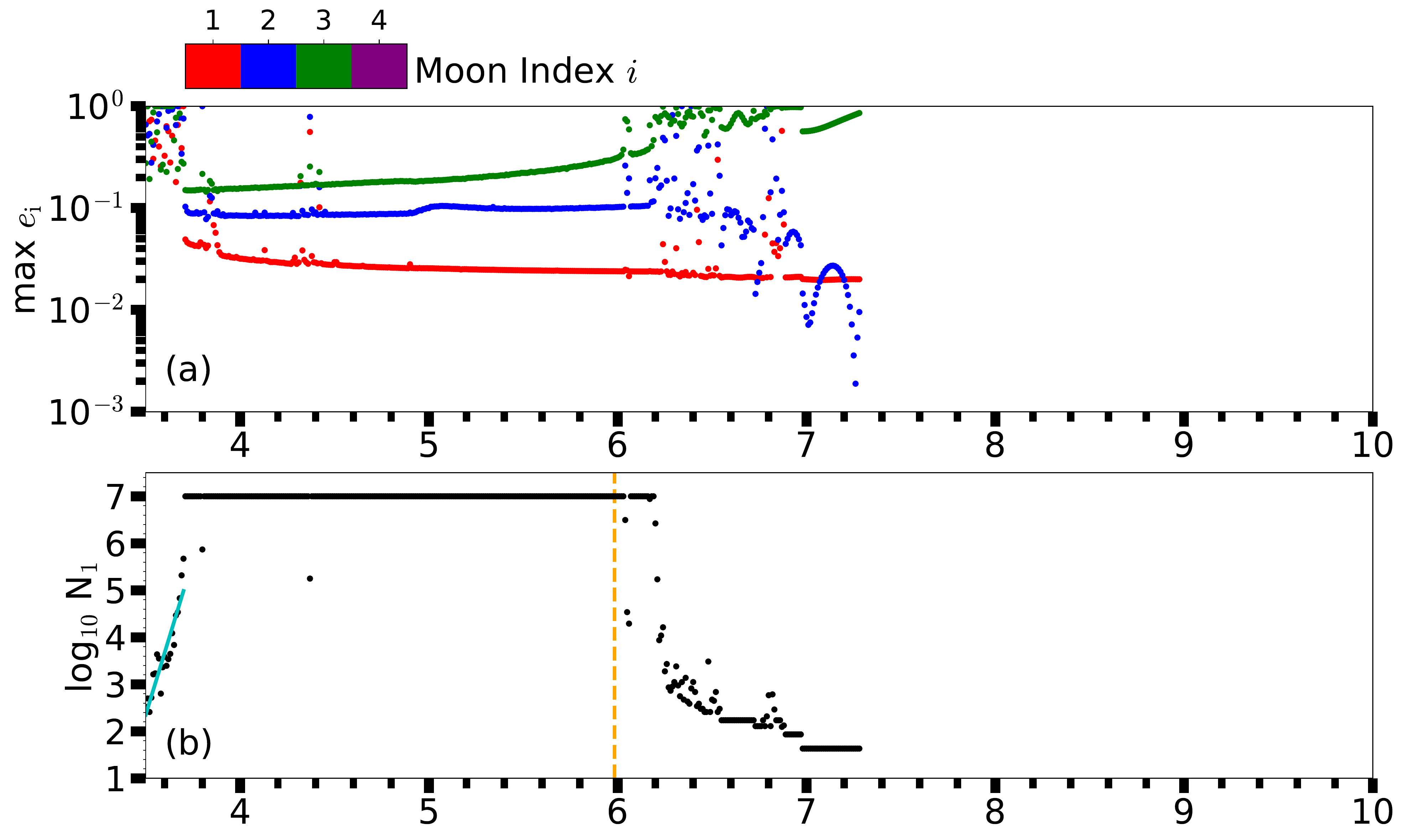}\\
  \includegraphics[width=0.7\linewidth]{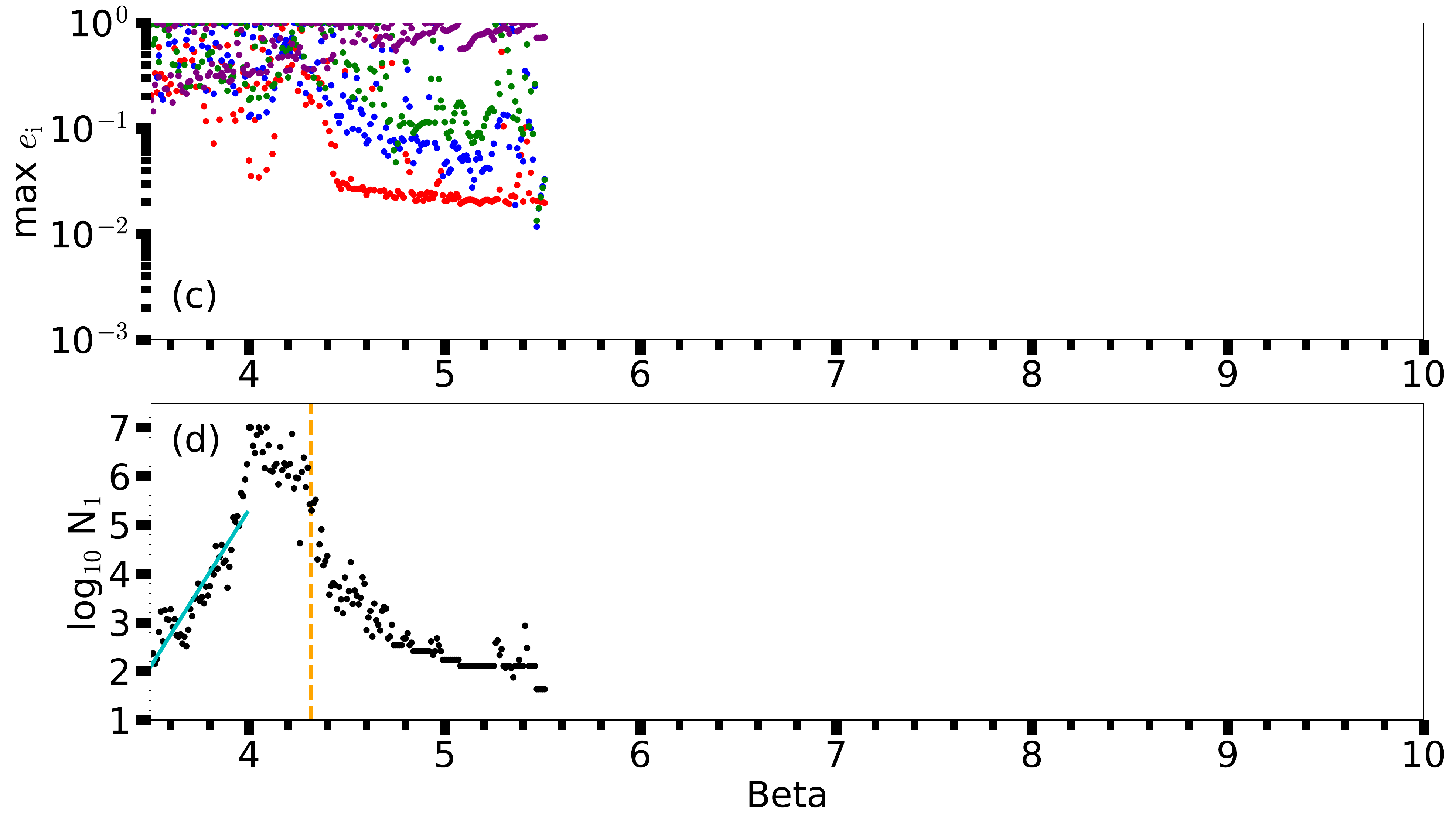}\\
\caption{Similar to Fig. \ref{fig:2} but for 3 (panels a \& b) and 4 (panels c \& d) Luna-Mass moons.}
\label{fig:4}
\end{figure*}

Table \ref{tab:beta_max} shows that the number of Luna-mass moons between the Roche radius and the Hill radius is limited to 4 due to requirement of $\beta_{min} = 2\sqrt{3}\approx 3.5$. Thus, we perform simulations considering only 3 and 4 Luna-mass moons following a similar procedure as in Sections \ref{ceres} and \ref{pluto}. The maximum eccentricity and system lifetime for three moons is shown in Figs. \ref{fig:4}a and \ref{fig:4}b, while Figs. \ref{fig:4}c and \ref{fig:4}d illustrate the same measures for four Luna-mass moons. Three moons can maintain stable orbits for a narrow range in spacing, $4< \beta < 6$. An additional moon (Figs. \ref{fig:4}c and \ref{fig:4}d), shows that the MMRs encountered by the outermost moon from the Sun and the overlap of MMRs between moons destabilizes the entire system.

\cite{Quarles2018} explored planet packing in $\alpha$ Centauri, where the secondary star significantly perturbs the planetary system and they showed that 3-planet systems can be long-lived with an appropriately chosen orbital spacing and initial eccentricity of the planets. The system architecture of planets orbiting one star of a stellar binary is similar to our system of moons orbiting an Earth-mass planet, where the forced eccentricity and boundaries of orbital stability limit the total number of satellites.  Due to the similarity in structure, we arrive at similar conclusions.

\subsection{Log-Linear fit of unstable lifetimes}
\label{sec:log-linear}

To compare the system lifetimes of different orbital architectures in planet packing, previous studies \citep{Chambers1996,Smith2009,Obertas2017,Quarles2018} employ a log-linear fit to the first transition region to stability $(\beta \sim4-8)$.  In figures \ref{fig:2}, \ref{fig:3}, and \ref{fig:4}, we mark these approximately linear trends using cyan lines, where we measure the slope $b$ and y-intercept $c$.  \cite{Quarles2018} showed that a constant $\beta$-shift to account $\beta_{\rm min}=2\sqrt{3}$ is necessary to ensure a fair comparison between simulated systems.  As a result, we fit the unstable data points in the transition region to a log-linear function of the form:

\begin{equation} \label{eqn:log_linear}
     \log_{10}\,t = b^\prime\beta^\prime + c^\prime,
\end{equation}

where the prime ($\prime$) coordinates refer to fits made with a shift in the $x$-axis (i.e., $\beta^\prime = \beta-2\sqrt3$).  Consequently, this shift in the x-axis minimizes the correlation between the slope ($b^\prime$) and the y-intercept ($c^\prime$).  This fit function is similar to the ones used by \cite{Quarles2018} and \citep{Lissauer2021} and has allowed us to make consistent comparison of the coefficients from the previous work.

Table \ref{tab:best_fit_coeff} shows a general trend in the coefficients of the log-linear function for all moon-types, where fewer moons $(n)$ correspond to a steeper slope $(b^\prime)$ and a longer system lifetime $(c^\prime)$ at $\beta_{\rm min}$.  The slope of the fit indicates how the system lifetime changes as a function of the orbital spacing parameter $\beta$.  For example, a steeper slope conveys that an increase in $\beta$ significantly extends the system lifetime.

For Ceres-mass moons, the slope only decreases by $\sim10\%$ as more moons are added.  The slopes for Ceres-mass moons are more similar to values determined through packed three planet systems around a single star \citep{Lissauer2021} rather than within a binary like $\alpha$ Centauri AB \citep{Quarles2018}. This is likely due to the minimal forcing from the Sun, especially when the moons occupy a smaller portion of the host planet's hill radius.  Table \ref{tab:beta_max} suggests that more Ceres-mass moons could stably orbit the host planet, but the moons are perturbed and scattered by the neighboring moons, (or the star) due to their lower inertia. Hence, the system of Ceres-mass moons is largely unstable for $\beta$ spacing $\le$ 6.5. The decreasing slopes also indicate that the orbital spacing $\beta$ must be increased, for increasing n, in order to avoid orbital crossings and maintain stability.

The other massive moons (Pluto-mass and Luna-mass) can absorb more internal (moon-moon) perturbation due to the increased inertia, or smaller changes to their angular momentum. Hence, the slopes for 3-moon cases are much steeper compared to the Ceres-mass(6 times for Pluto-mass and 12 times for Luna-mass). In addition, pairs of more massive moons have a wider mutual Hill radius which reduces the number of moons that can fit within the stability boundary. Therefore, the number of stable moons for Ceres-mass is $\le$ 8, for Pluto-mass is $\le$ 5 and for the Luna-mass is $\le$ 4. The decreasing slopes for Pluto-mass and Luna-mass moons enforces our previous conclusion that the $\beta$ must be increased, for increasing n, to avoid the orbital crossings.

\begin{table}
	\centering
	\caption{Coefficients for the log-linear fits (Cyan colored lines in Figs. \ref{fig:2}-\ref{fig:4}) using $\log_{10}(t) = b^\prime \beta^\prime + c^\prime$ (Eqn. \ref{eqn:log_linear}). The primed values ($m^\prime$ and $b^\prime$) constitute the shift in $\beta$ by $2\sqrt3$. $^1$ A system of up to five Earth-mass planets orbiting $\alpha$ Cen B \citep{Quarles2018}. $^2$ A hypothetical system, where three Earth-like planets orbit a Sun-like star \citep{Lissauer2021}.}
	\label{tab:best_fit_coeff}
	\begin{tabular}{lccc}
\noalign{\smallskip}
\hline
\noalign{\smallskip}
Mass & n & Slope & y-intercept \\
           &  &  ($b^\prime$) & ($c^\prime$) \\
\noalign{\smallskip}
\hline
\noalign{\smallskip}
Ceres & 3 & 1.37 & 2.21 \\
    & 4 & 1.32 & 1.83 \\
    & 5 & 1.31 & 1.59 \\
    & 6 & 1.25 & 1.56 \\
    & 7 & 1.24 & 1.53 \\
    & 8 & 1.24 & 1.51 \\
\hline
Pluto & 3 & 6.45 & 0.50 \\
    & 4 & 5.35 & 0.24 \\
    & 5 & 5.01 & 0.20 \\
\hline
Luna & 3 & 12.44 & 1.96 \\
    & 4 & 6.4 & 1.88 \\
\hline
Earth-mass$^1$ & 3 & 0.996 & 2.234 \\
               & 5 & 0.742 & 2.084 \\
\hline
Earth-mass$^2$ & 3 & 1.68 & 1.799 \\
\noalign{\smallskip}
\hline
\noalign{\smallskip}
\end{tabular}
\end{table}

\subsection{Analysis of MMR using MEGNO maps}
\label{mmr_megno}

We explore the potential routes to instability in systems of multiple moons by using the chaos indicator MEGNO (mean exponential growth of nearby orbits $\langle Y \rangle$) maps. The MEGNO criterion is generally used to distinguish between chaotic, periodic, and aperiodic orbits within a phase space. Analysis of many body systems using MEGNO was originally developed by \cite{Cincotta1999,Cincotta2000,Cincotta2003} to identify potential instabilities due to resonance overlap more efficiently in a short numerical integration period. It is capable of detecting high-order resonances (for example, see \cite{satyal2014} for a MEGNO map of a circumprimary planet displaying 39:2 MMR in a binary system) due to its sensitivity to unstable orbits and is a global indicator of dynamical changes in any Hamiltonian system. In our case, we are exploiting this particular nature of MEGNO which can reveal fine resonance structures in a phase space to display the unstable and chaotic regions.

We have limited the phase space up to $e_o$ = 0.3 because (for higher $e_o$) it is mostly chaotic region induced by the secular evolution of the eccentricities, see for example Fig. 8 from \cite{Tamayo2021}.

\begin{figure*}
  \includegraphics[width=0.7\linewidth]{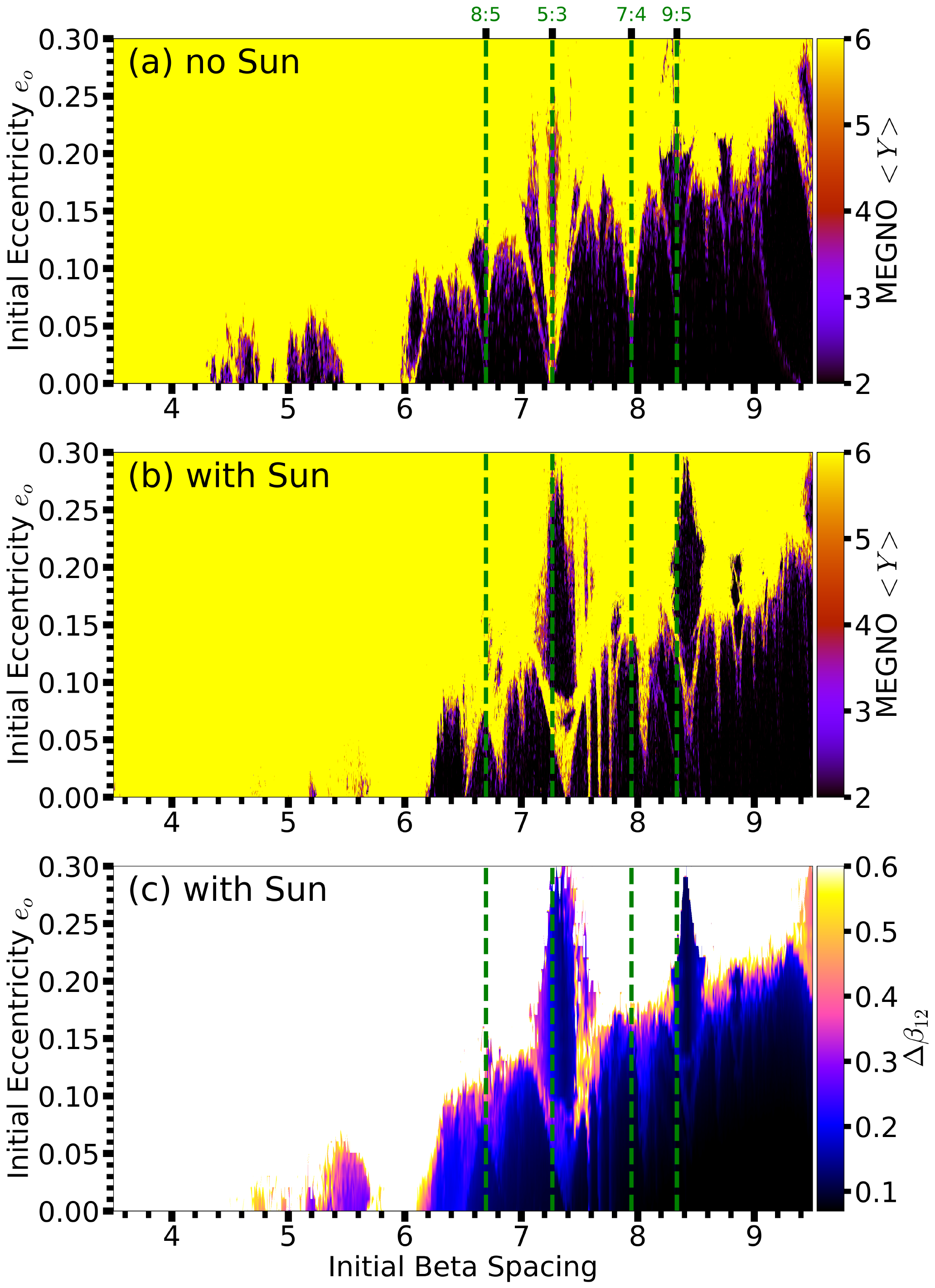}
\caption{MEGNO map for a system with three Ceres-mass simulated for $\sim $2 million orbits of the innermost moon. Panels (a) and (b) show the measure of MEGNO $\langle Y \rangle$ for the modified system (i.e., Earth and moons only) and full system (i.e., Sun-Earth-moons), respectively. MEGNO $\langle Y \rangle$ values equal to 2 indicate periodic orbits while higher values indicate chaotic and potentially unstable orbits.  Panel (c) shows the variation in $\beta$ between the first 2 moons that is attained over a simulation. The green vertical lines represent the estimated locations of the MMRs \citep{Murray2000}. The MMR offset between the expected locations and observed numerical solutions shows the influence of the Sun on the satellite systems.}
\label{fig:5}
\end{figure*}

Figure \ref{fig:5}a displays MEGNO values (color-coded)  calculated for the outermost moon in a system with an Earth-mass planet and three Ceres-mass moons (ignoring the Sun) with a wide variation in the initial $\beta$ spacing and eccentricity (of the outermost moon) in the $e_o$ - $\beta$ phase space. MEGNO values where $\langle Y \rangle$ = 2 corresponds to periodic (and presumably stable) orbits, where $\langle Y \rangle$ = 6 indicates initial conditions that drive the system into chaos. Other quasi-periodic orbits can exist between these extremes (purple-orange). The outermost moon's initial eccentricity is varied from 0 to 0.3 and the spacing $\beta$ between all the moons ranges from 3.5 to 9.5. Most regions in the chaotic regions of the phase space (yellow) are  unstable, where the outermost moon has a high probability of collision with the host planet or another one of the moons.  The existence of the chaotic regions are largely due to MMR overlap or secular evolution of the eccentricity \citep[i.e.,][]{Wisdom1980,Mudryk2006,Quillen2011,Laskar2017,Hadden2018,Petit2020,Tamayo2021}. The stable orbits (black) begin when $\beta \gtrsim 6$, and for lower initial eccentricity. For one initial condition ($e_o$ = 0), the map compliments the analysis presented in Sec. \ref{ceres}, Fig. \ref{fig:2}a, where the orbits are stable for the full integration period when a higher $\beta$ spacing is selected. Figures \ref{fig:5}b and \ref{fig:5}c both explore a similar phase space, but now including the Sun in the simulations.

Multiple MMRs are V-shaped (because their libration width increases with increasing eccentricity) in the map, where the 8:5, 5:3, 7:4, and 9:5 are some of the prominent MMRs between moons and are marked (dashed green lines). Figure \ref{fig:5}b demonstrates how the Solar perturbations affect which of the initial conditions are affected by the moon-moon MMRs.  As a result, there is an offset observed in the MEGNO phase-space when comparing Figs. \ref{fig:5}a and \ref{fig:5}b. For example, Fig. \ref{fig:5}a shows the  5:3 MMR located at $\beta$ = 7.2, but the MMR structure is located at $\beta$ = 7.4 in Fig. \ref{fig:5}b. This shift is accounted for by the secular perturbations on the moons by the Sun, which causes the orbital spacing of the moons to change over time.

A moon can then evolve into and out of a nearby MMR.  Figure \ref{fig:5}c illustrates the shifts in orbital spacing between the two inner moons through $\Delta\beta_{12}$, which signifies the maximum change in the orbital spacing.  In and around the observed MMRs, the shift in the $\Delta\beta$ is about 0.1, which accounts for the differences in the location of MMRs in Figs. \ref{fig:5}a and \ref{fig:5}b . Also observed in Fig. \ref{fig:5}c is that the $\Delta\beta$ values bigger than 0.3 indicates unstable orbits and those values that are less than 0.3 are stable. It is also apparent from the map that the orbits with $\Delta\beta$ less than 0.2 are not heavily influenced by any MMRs.

\begin{figure*}
  \includegraphics[width=0.7\linewidth]{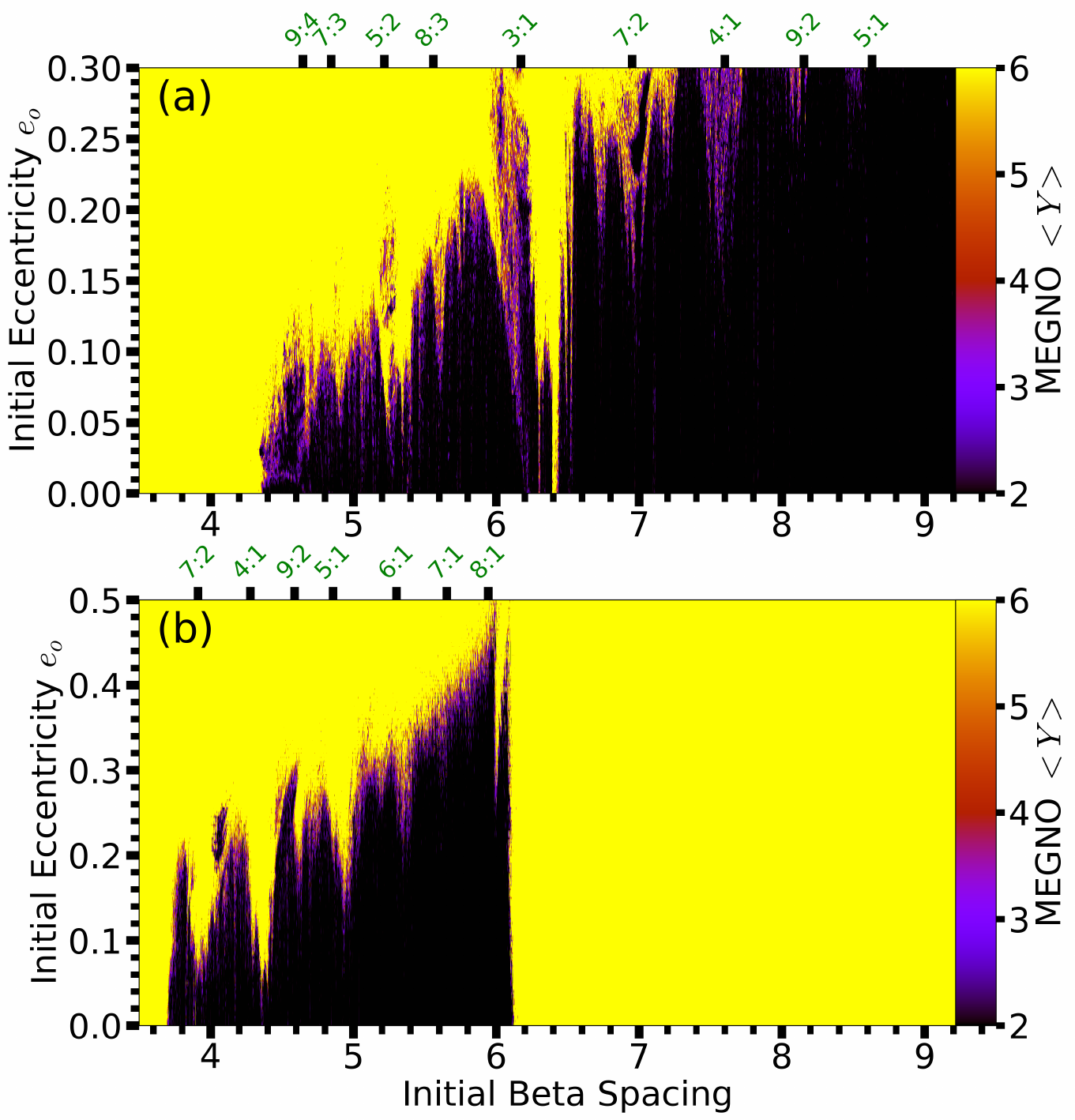}
\caption{Similar to Fig. \ref{fig:5}b, but for a system with three (a) Pluto-mass and (b) Luna-mass. }
\label{fig:6}
\end{figure*}

A similar phase space is explored for larger moons, but focusing on three moon systems with both the host planet and the Sun included in Fig. \ref{fig:6}a (Pluto-mass moons) and \ref{fig:6}b (Luna-mass moons).  Figure \ref{fig:6}a shows that the outermost Pluto-mass moon can start with relatively large eccentricity as long as the orbital spacing is large enough $(\beta\gtrsim 6.5)$, while an equivalent system with Luna-mass moons is more limited in terms of its initial orbital spacing (Fig. \ref{fig:6}b).  Both systems require larger exchanges of angular momentum to perturb their orbits, as compared to the Ceres-moon case (Fig. \ref{fig:5}b), which explains why a higher initial eccentricity can allow for stable, periodic orbits (black regions).  Pluto-mass moons appear the most optimal as nearly half of the parameter space allows for periodic orbits.  However the MMR at $\beta \sim 6.4$ may induce instabilities as a primordial moon system evolves outward due to tidal interactions with the host planet. Also, at this resonance, the moons'  e$_{max}$ are observed to evolve towards 1, Fig. \ref{fig:3}a.

\subsection{Shifting the MMRs}
\label{Orb_param}
\begin{figure*}
   \includegraphics[scale=.22]{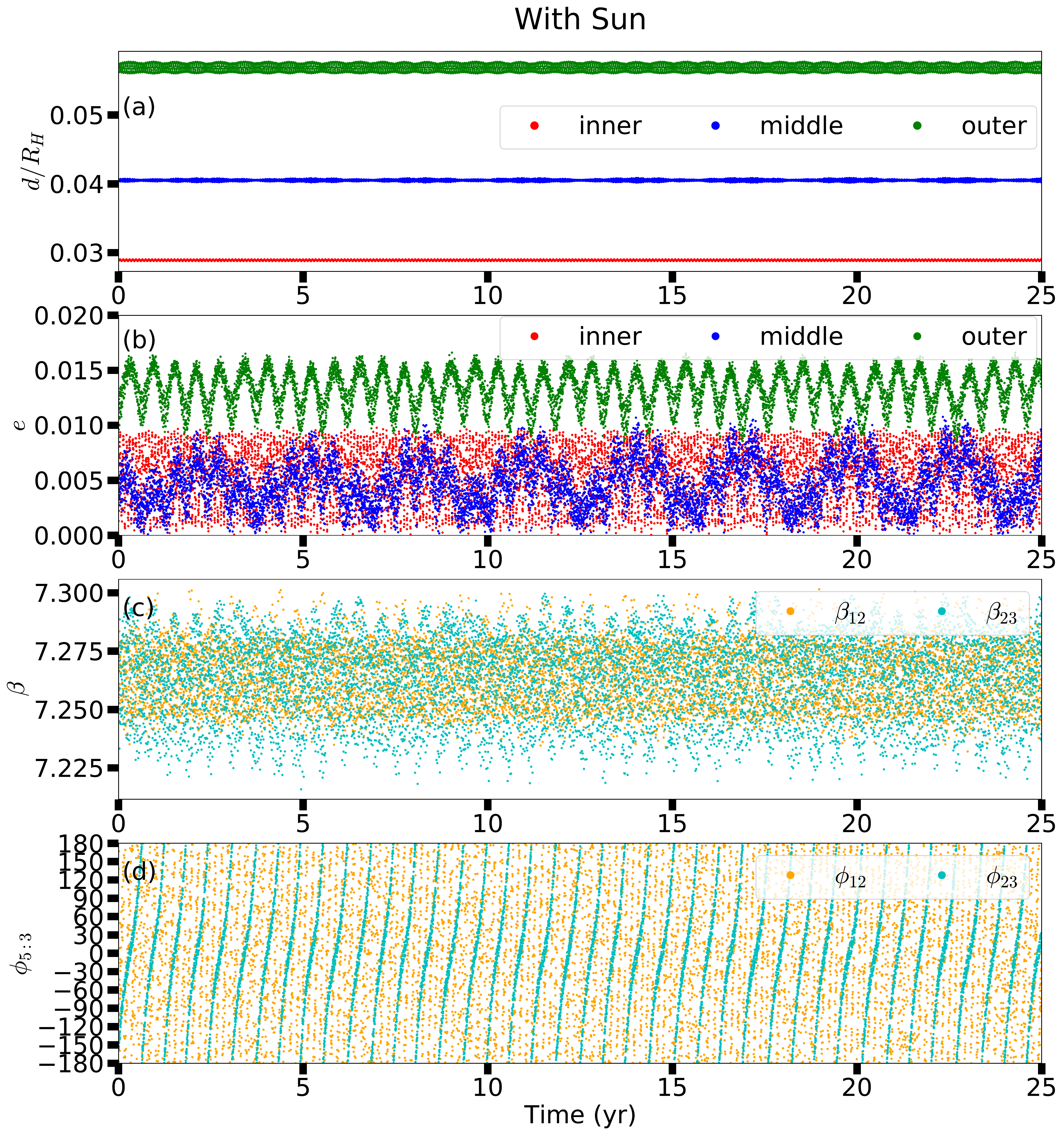}
  \includegraphics[scale=.22]{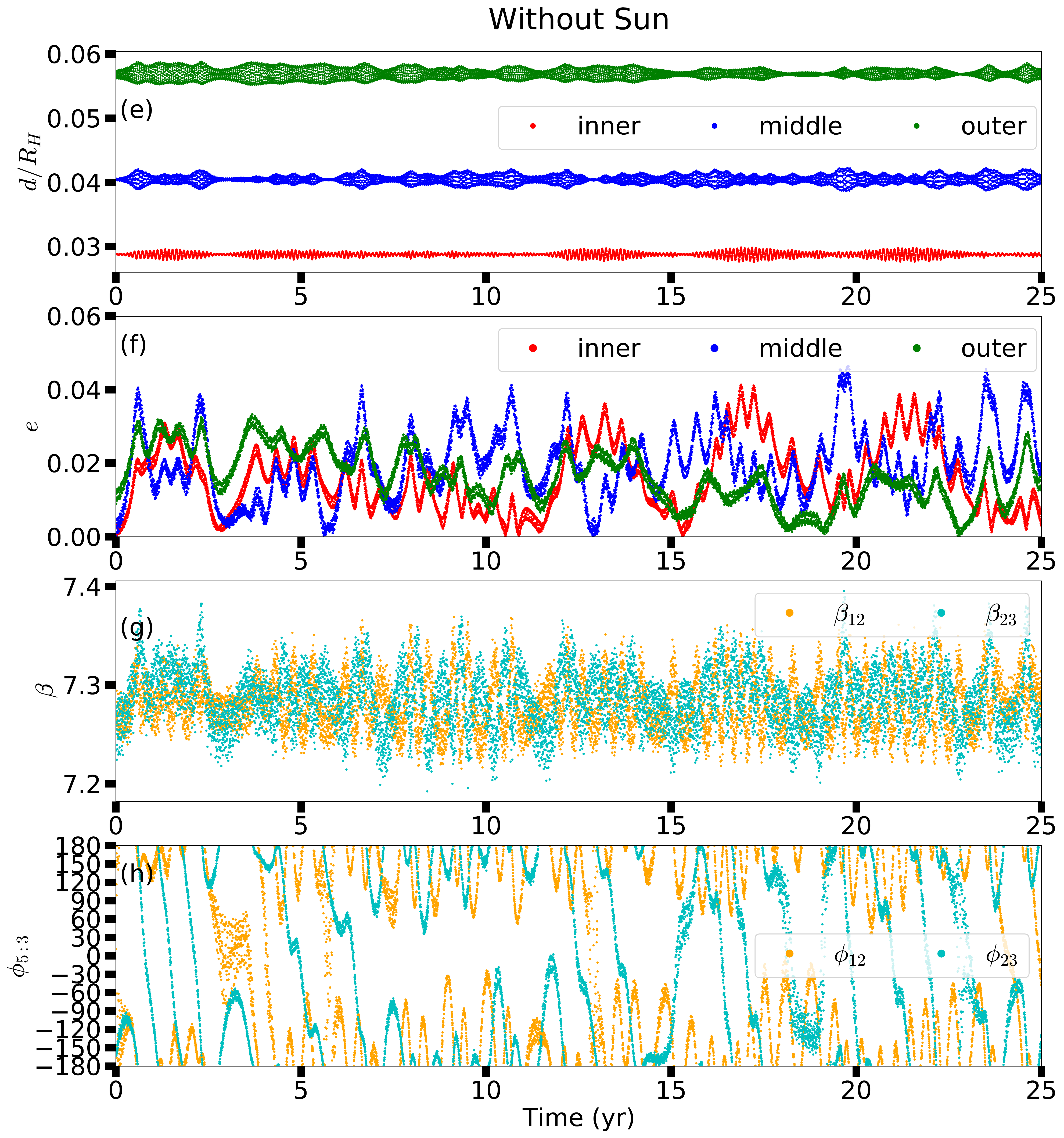}
 \caption{Orbital evolution of three Ceres-mass moons with respect to their scaled distance from the host planet $d/R_H$ (panels a \& e), eccentricity $e$ (panels b \& f), orbital spacing $\beta$ (panels c \& g), and the 5:3 resonant angle $\phi$ between pairs of moons (panels d \& h). The systems are simulated for 25 years with an initial orbital spacing $\beta=7.26$ (i.e., the resonance location for the 5:3 MMR in Fig. \ref{fig:6}). The orbital evolution on the left includes the Sun in the N-body simulation, where the panels on the right exclude the Sun.}
 \label{fig:7}
 \end{figure*}

The inclusion (or exclusion) of the Sun and its effect in the dynamics of the moons are viewed in global phase space maps ($e_o$ vs. $\beta$) using by the MEGNO criterion (Fig. \ref{fig:5}). For the case with 3 Ceres-mass moons, the 5:3 MMR is clearly observed at $\beta$ = 7.26 without the Sun, Fig. \ref{fig:5}a) and slightly shifted higher at $\beta$ = 7.4 (with the Sun, Fig. \ref{fig:5}b). To visualize the time-series data of individual orbital elements we use initial conditions ($\beta=7.26$ and $e_o$ = 0.01) from Fig. \ref{fig:5} and simulate the system for 25 years ($\sim 3\times 10^6$ orbits of the innermost moon).  Figure \ref{fig:7} shows the time-series evolution of the normalized orbital distance $d/R_H$, eccentricity $e$, orbital spacing $\beta$, and resonant angle $\phi_{5:3}$ for each moon in both systems (with and without including the Sun). Note that the two (inner and middle) moons begin on circular orbits, while the outer moon begins with the eccentricity $e_o$ = 0.01, to maintain consistency with Fig. \ref{fig:5}. The normalized distance ($d/R_H$) is used instead of semimajor axis to better illustrate the correlated changes in distance with eccentricity as it affects the instantaneous measure of the orbital spacing.

In Fig. \ref{fig:5}b ($\beta=7.26$ and $e_o$ = 0.01 with the Sun included), the value of MEGNO suggests that the orbits are periodic. This is confirmed in the evolution of each moon's normalized distance (Fig. \ref{fig:7}a) and eccentricity (Fig. \ref{fig:7}b, where the colors refer to the inner (red), middle (blue) and outer (green) moons. The gravitational perturbation of the Sun forces a high frequency variation in the relative distance of each moon, which underlies the variation in the orbital spacing $\beta$ (Fig. \ref{fig:7}c.  This forcing also prevents the 5:3 MMR resonant angle of the inner pair $(\phi_{12})$ or the outer pair $(\phi_{23})$ of moons from librating (in Fig. \ref{fig:7}d), even though the respective orbital spacing of each pair ($\beta_{12}$ in orange, or $\beta_{23}$ in cyan) crosses the expected MMR location $\beta = 7.26$.  
In contrast, Figs. \ref{fig:7}e-\ref{fig:7}h display a similar simulation, where the Sun is removed (i.e., ignoring its secular perturbation).  In this case, the normalized distance and the eccentricity of each moon are chaotic, which is indicated by the corresponding MEGNO value from Fig. \ref{fig:5}b.  The variation of each moon's eccentricity (in Fig. \ref{fig:7}f) can be $2-4$ times larger as compared to Fig. \ref{fig:7}b due to the eccentricity excitation from the 5:3 MMR.  The orbital spacing of the inner and outer moon pairs vary on much slower timescales in Fig. \ref{fig:7}g, which allows for moon pairs to evolve together into the 5:3 MMR with an average $\beta \sim 7.26$.  The switching between libration and circulation in Fig. \ref{fig:7}h confirms the chaotic nature of this initial condition and the connection with the 5:3 MMR.

For an Earth-mass planet with a semimajor axis of 1 AU, the Sun clearly contributes to the dynamics of the planet and its moons. The inclusion of the Sun in a system  is necessary for our analysis.  However, the chaotic dynamics that remains by removing the Sun is still applicable, where the relevant initial orbital spacing $\beta$ must be shifted by $\sim0.15$ to $\beta \sim 7.4$.  Alternatively, moon systems of long-period Earth-mass planets could undergo similar chaotic variations (e.g., Pluto-Charon and its four moons).

\subsection{Mass Distribution and Formation Plausibility}
\label{surface_density}

Based on the number of moons that stably orbit the planet, we calculate the total mass distribution within the stability boundary for all three systems.  The total mass of 3 Luna-mass, 5 Pluto-mass, and 8 Ceres-mass moons is equivalent to 1.11 x 10$^{-7}$ M$_\odot$, 3.30 x 10$^{-8}$ M$_\odot$, and 3.60 x 10$^{-9}$ M$_\odot$, respectively.  Despite the small differences in the semimajor axis of the innermost moon $a_1$, the order of magnitude differences in mass (see Table \ref{tab:Orb_Param}) are preserved.  Since the mass is distributed roughly in the same surface area between the inner (i.e., Roche limit) and outer stability boundaries, the surface mass density for Ceres ($\sigma_{Ceres}$) is approximately 7.2 x 10$^{-5}$ M$_\odot$/AU$^2$, equivalent to 637 g/cm$^2$. The surface mass densities for Pluto and Luna are 10$\sigma_{Ceres}$, and 30$\sigma_{Ceres}$, respectively. 

Our work concentrates on the maximum number of satellites, with different masses, that can stably orbit around an Earth-like planet. Whether more than one satellite can form around an Earth-like planet is beyond the scope of this work.  However, we can provide some assessment on the plausibility of formation based upon our results of the total mass surface density for the maximum number of moons. Moons can form around a giant planet (e.g., Jupiter and Saturn) from a gaseous circumplanetary disk during the last stages of planet formation as has been shown for the Galilean moons \citep{pollack1989,Canup2006}. But, there is not a known minimum planetary mass threshold at which a circumplanetary disk can form and evolve into moons \citep{Ayliffe2009}. Large moons are expected to arise, primarily due to giant impacts, where \cite{Nakajima2022} suggests that impact-induced large moons are more likely to form around rocky planets whose radius is smaller than 1.6 R$_\oplus$. Moreover, numerical models using smooth particle hydrodynamics (SPH) have shown that the mass surface density of moon-forming disks can reach $\sim 10^7$ g/cm$^2$ only at a few Earth radii and substantially spread over the lifetime of the disk \citep{Nakajima2014,Nakajima2022}. Therefore, it is at least plausible that multiple moons could  form around Earth-like planets.  Further study or confirmation of the current exomoon candidates (e.g., Kepler 1625b-i \cite{Teachey2018}; Kepler 1708b-i \cite{Kipping2022}) could shed more light on these hypotheses.

\subsection{Effects of Tides}
\label{section:tides} 

The outward migration of satellites through tidal interactions modifies their potential lifetime \citep{Barnes2002, Sucerquia2019, Lainey2020}. In the Solar System, \citet{Charnoz2010} suggested that the population of small moons that orbit just outside Saturn's rings could have originated at the edge of the main rings and tidally migrated outward. To obtain a complete picture of the orbital stability of exomoons, it is necessary to consider the contribution of planetary and stellar tides. We apply a secular constant time lag (CTL) tidal model \citep{Leconte2010, Hut1981} and evaluate the migration timescales of moons assuming that moon formation readily occurs near the host planet's Roche limit.  The lifetime of a moon system can be reduced as the outermost moon migrates toward the outer stability limit (i.e., $\sim 0.40 R_H$; \citet{yagirl2020}), where this will depend on the mass of the satellite (or moon-planet mass ratio) and the assumed time lag $\Delta t$ for the tidal dissipation. The secular model calculates the changes to the orbital elements of both the host planet and its moon through the respective semimajor axes ($a_{\rm p}$ and $a_{\rm sat}$), eccentricities ($e_{\rm p}$ and $e_{\rm sat}$), and mean motion ($n_{\rm p}$ and $n_{\rm sat}$) averaged over an orbit.  The model is scaled by the tidal Love number $k_2$ and the time lag $\Delta t$, where the latter is proportional to $(nQ)^{-1}$ in the constant phase lag (i.e., constant $Q$) tidal models \citep{Leconte2010, Piro2018}.

We consider two scenarios for three moon systems: a) keep the satellite mass fixed at $3\ m_{Luna}$, while evaluating a range of constant time lag values, and b) keep the time lag fixed at $\Delta t = 100\ {\rm s}$, while evaluating a range in satellite mass (Ceres-, Pluto-, and Luna-mass).  The constant time lag $\Delta t$ is varied between simulations from $10-600\ {\rm s}$ on a logarithmic scale.  The moon-moon interactions are ignored since we are applying a secular model, where the total mass of the moon system is combined into the innermost moon.  This represents a conservative estimate because the innermost moon would be migrating outward more slowly in reality as it would be 1/3 of our prescribed mass. The Earth-mass host planet is assumed to begin with a rotation period of 5 hours, which is consistent with expectations from terrestrial planet formation \citep{Kokubo2007}, and we are interested in a $10^{10}$ year timescale (i.e., the main-sequence lifetime of a G dwarf). 

\begin{figure*}
	\includegraphics[width=.7\linewidth]{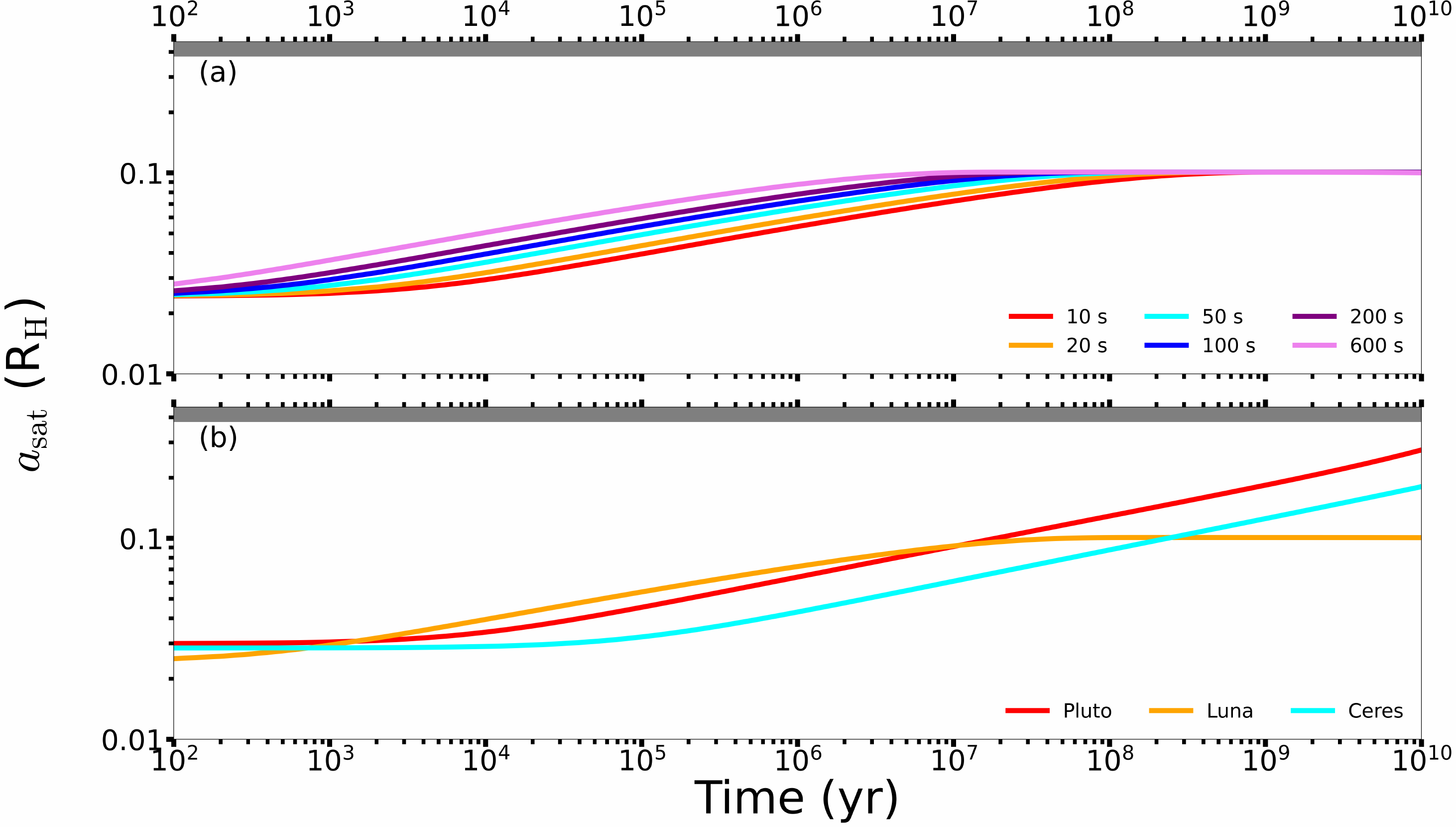}
    \caption{Outward migration within a constant time lag (CTL) tidal secular model of a Sun-Earth-moon system that varies the (a) time lag or (b) assumed satellite mass.  In panel (a), the satellite mass is $3\ m_{Luna}$ while the time lag varies (color-coded).  In panel (b), the satellite mass is varied ($3\times$ Ceres-, Pluto-, and Luna-mass) and the time lag $\Delta t$ is 100 s. The gray region marks the unstable region, past the stability limit, for prograde orbits.  }
    \label{fig:tides_secular}
\end{figure*}

In our first scenario, shown in Figure \ref{fig:tides_secular}a, the mass of the satellite is $3\ m_{Luna}$ and the constant time lag is varied over a range that corresponds to a very low dissipation (10 s; red) up to a very high dissipation (600 s; lavender).  In all cases, the satellite's outward migration stalls at $\sim 0.1\ R_H$ as the moon's orbital period synchronizes with the planet's rotation period.  Assuming that the moons migrate outward together maintaining an orbital spacing of $\beta = 4$, then the outermost moon would migrate beyond the outer stability limit (i.e., $a_3 \sim 0.55\ R_H$) and thereby reduce the number of stable moons by one.

The effect of tidal migration is more dire for lower mass moons because they can migrate closer to the outer stability limit (Fig. \ref{fig:tides_secular}b) and require a larger orbital spacing (see Figs. \ref{fig:5}b and \ref{fig:6}a).  This combination of circumstances will likely cause at least 1 moon to scatter and/or migrate past the outer stability limit.  Therefore, outward tidal migration will likely reduce the number of moons orbiting an Earth-mass planet by at least one in three moon systems and likely more within moons systems of higher multiplicity. Further consideration of tides is beyond the scope of our work, where others could explore the effects of differential tides on outward migration that is similar to models for Saturn's moons \citep{Crida2012,Cuk2016}. 

\section{Conclusions}
\label{section:conclusions}
Through n-body simulations, we investigate the potential for systems of $3-9$ moons orbiting an Earth-mass planet and a Solar-mass star.  The moons vary in mass, but are analogous to Ceres, Pluto, and our Moon (Luna).  Systems of multiple moons are inherently constrained by the \emph{inner} Roche limit and the \emph{outer} stability limit, which can be also scaled by a planet's Hill radius.  Scaling by the Hill radius allows our work to be generalized beyond an exact Earth-Sun analog for the primary bodies, because the Hill radius incorporates potential changes in the planetary semimajor axis, eccentricity, and mass, in addition to the stellar mass.  Each moon system begins on a circular and coplanar orbit, where the initial orbital phase is selected through the golden ratio following planet-packing studies \citep{Smith2009,Quarles2018,Lissauer2021}.  We find using N-body simulations that $7\pm1$ Ceres-mass moons could stably orbit an Earth-mass planet at 1 AU from a Sun-like star. If the moons are more massive (Pluto- or Luna-mass), then the number of moons with stable orbits reduces to $4\pm1$ and $3\pm1$, respectively.  Outward tidal migration will likely modify these estimates by at least one moon, where additional moons could be lost through scattering, collisions, or simply migrating beyond the outer stability limit.

The orbital spacing between each moon is measured using a dimensionless parameter $\beta$, which is the distance between two neighboring moons divided by their mutual Hill radius.  The maximum number of moons produces a minimum in the orbital spacing, where we find a $\beta = 6,\ 4.5,\ \text{and}\ 3.5$ for Ceres-, Pluto-, and Luna-mass moons, respectively.  The potential stability for these moon systems depends on their proximity to MMRs between adjacent pairs of moons.  The location of the MMRs are estimated using the chaos indicator MEGNO, which shows a shift of $\sim0.15$ in $\beta$ from the expected location due to perturbations on the moon system from the Sun.  We show that a moon can behave chaotically (i.e., periodic switching between circulation and libration in the 5:3 resonant angle) when starting within the libration zone of the 5:3 MMR and ignoring the gravitational solar perturbations.  

Planet-packing studies have used a best-fit slope from a log-linear model to compare the changes in stability due to planet multiplicity \citep{Chambers1996,Smith2009,Pu2015,Obertas2017,Quarles2018}.  We employ a similar technique, but in shifted coordinates so that the $y$-intercept occurs at $\beta = 2\sqrt{3}$ \citep[e.g.,][]{Quarles2018,Lissauer2021}.  From these measurements, we find that Ceres-mass moons have a slope from $1.24-1.37$, which is inversely correlated with the number of moons.  These slopes are less than the expected values for systems of three Earth-mass planets orbiting a Solar-mass star and greater than the more extreme case where three planets are orbiting $\alpha$ Centauri B, as stellar binary.  Pluto- and Luna-mass moons have a much steeper slope because they have a larger mutual Hill radius, which drastically limits the potential orbital spacing between moons (see Table \ref{tab:beta_max}).

We use the mass surface density within the stability boundary limits to determine whether systems of multiple moons are at least plausible.  \cite{Nakajima2022} showed that the surface density of a moon-forming disk can reach $\sim 10^7$ g/cm$^2$, which is much higher than the mass surface density of three Luna-mass moons ($\sim 10^4$ g/cm$^2$).  The mass surface density for five Pluto-mass moons is smaller by a factor of 3, while it decreases by a factor of 30 for eight Ceres-mass moons.  It appears plausible that multiple moons could form, but further study using SPH simulations would be necessary and is beyond the scope of our work.

Our N-body simulations are mostly limited to only $10^7$ orbits of the innermost moon, which is ${\sim}3\times 10^4$ yr.  The long-term evolution of moon systems will be determined by the outward migration of the moons due to tidal interactions with the host planet.  We evaluate this possibility using a constant time lag (CTL) secular model \citep{Hut1981,Leconte2010,Quarles2021}, where outward tidal migration becomes significant at long timescales ($\sim 10^8-10^{10}$).  As a result, the number of moons that can stably orbit an Earth-mass planet is reduced by one.  In the case of Luna-mass moons, only one moon is lost, where systems of less massive moons can have more significant losses due to the relative ease of scattering events and the final migration distance of the innermost moon.

Detecting multiple moon systems orbiting other stars is currently out of reach, where there are only a couple of exomoon candidates using the photometric detection method \citep{Teachey2018,Kipping2022}.  Recent observations from ALMA \citep{Benisty2021} suggest that a moon-forming disk exists around PDS 70c, which points to the potential for long-wave observations or direct imaging \citep{Vanderburg2018}.  From this work, the dynamical stability of moon systems limits the existence of exomoons for Earth-analogs in their respective habitable zones, where confirmation from future observations are needed.

\section*{Acknowledgements}
The authors appreciate the constructive comments and feedback from the referee.  M.R.F. acknowledges support from the NRAO Gr\"{o}te Reber Fellowship and the Louis Stokes Alliance for Minority Participation Bridge Program at the University of Texas at Arlington. This research was supported in part through research cyberinfrastructure resources and services provided by the Partnership for an Advanced Computing Environment (PACE) at the Georgia Institute of Technology.

\section*{Data Availability}
The data underlying this article will be shared on reasonable request to the corresponding author.



\bibliographystyle{mnras}
\bibliography{References} 


\bsp	
\label{lastpage}
\end{document}